\documentclass[12pt]{article}

\usepackage{authblk}
\usepackage{graphicx}
\usepackage[format=plain,font=it]{caption} 
\usepackage{booktabs}
\usepackage{amsfonts} 
\usepackage{amsmath} 
\usepackage{soul} 
\usepackage{multirow} 
\usepackage{subcaption}
\usepackage{hyperref}
\usepackage{adjustbox}

\usepackage[dvipsnames]{xcolor}

\title{Gaussian Process Regression models for the properties of micro-tearing modes in spherical tokamaks}

\author[1,\footnote{Corresponding author, email: william.hornsby@ukaea.uk}]{W.A Hornsby}
\author[1]{A. Gray} 
\author[1]{J. Buchanan}
\author[1]{B.S. Patel} 
\author[1]{D. Kennedy}
\author[1]{F.J. Casson} 
\author[1]{C.M. Roach} 
\author[1]{M. B. Lykkegaard} 
\author[2]{H. Nguyen}
\author[2]{N. Papadimas} 
\author[2]{B. Fourcin} 
\author[2]{J. Hart}

\affil[1]{UKAEA-CCFE, Culham Science Centre, OX14 3DB Abingdon, United Kingdom}
\affil[2]{digiLab, The Quay, Exeter, EX2 4AN, United Kingdom}
\date{\today}

\begin{document}
\maketitle
\begin{abstract}
    
Spherical tokamaks (STs) have many desirable features that make them an attractive choice for a future fusion power plant. Power plant viability is intrinsically related to plasma heat and particle confinement and this is often determined by the level of micro-instability driven turbulence. Accurate calculation of the properties of turbulent micro-instabilities is therefore critical for tokamak design, however, the evaluation of these properties is computationally expensive. The considerable number of geometric and thermodynamic parameters and the high resolutions required to accurately resolve these instabilities makes repeated use of direct numerical simulations in integrated modelling workflows extremely computationally challenging and creates the need for fast, accurate, reduced-order models.

This paper outlines the development of a data-driven reduced-order model, often termed a {\it surrogate model} for the properties of micro-tearing modes (MTMs) across a spherical tokamak reactor-relevant parameter space utilising Gaussian Process Regression (GPR) and classification; techniques from machine learning. These two components are used in an active learning loop to maximise the efficiency of data acquisition thus minimising computational cost. The high-fidelity gyrokinetic code GS2 is used to calculate the linear properties of the MTMs: the mode growth rate, frequency and normalised electron heat flux; core components of a quasi-linear transport model. Data cross-validation and direct validation on unseen data is used to ascertain the performance of the resulting surrogate models.

\end{abstract}

\section{Introduction}
\label{sec:intro}

Spherical tokamaks possess a number of potential advantages when compared to conventional aspect ratio devices. These include improved vertical stability properties as well as the ability to access high $\beta$ regimes of operation, allowing a more compact and ultimately cheaper reactor design. However, in these scenarios electromagnetic instabilities, such as kinetic ballooning modes (KBM) \cite{Tang_1980} and micro-tearing modes (MTM) \cite{hazeltine1975}, are typically excited, both of which can drive radial heat propagation away from the plasma core and potentially degrade plasma performance.
 
In the case of MTMs, the turbulence distorts and reconnects the equilibrium magnetic field, resulting in microscopic magnetic island structures which can overlap and cause field stochasticity, which is believed to be responsible for large radial electron heat transport\cite{rechester1978,guttenfelder2011,guttenfelder2012,Giacomin_2023a}. Consequently, being able to reliably predict the critical gradients at which MTMs are expected to be driven unstable has considerable value for the design of performant reactor scenarios.

The most appropriate framework for describing turbulent modes in the core of a tokamak is the gyrokinetic framework \cite{Sugama_2000,PJCatto_1981}. Running non-linear gyrokinetic codes routinely within integrated models for design space exploration and turbulence modelling is very costly, particularly for electromagnetic modes like MTMs which exhibit extremely fine radial structure and consequently require very high numerical resolution. Even local linear gyrokinetic simulations are currently prohibitively expensive in the context of integrated modelling over confinement timescales, being many orders of magnitude slower than currently used reduced turbulence models such as TGLF.

An alternative to repeated direct use of high-fidelity models in integrated workflows is to use a training dataset to build a data-driven reduced-order model of the simulation outputs using machine learning based techniques. This is an especially attractive approach in the case of MTMs due to the lack of sufficiently accurate physics-based reduced-order models valid for the highly shaped plasmas found in STs, however progress is being made in the development of valid reduced models \cite{staebler_23,Avdeeva_2023}.

Machine learning (ML) and uncertainty quantification (UQ) techniques have already been extensively used in transport and turbulence modelling \cite{Citrin_2015,Pavone_2023,farcas_2022a,Citrin_2023,DASBACH2023101396,vandePlassche_2020}, including with fully nonlinear plasma turbulence simulations in a limited way \cite{rodriguez_fernandez_2018,DiSiena_2022}, to create predictive surrogate models. Furthermore neural networks trained on experimental databases\cite{QualikizNN_21} and on reduced micro-tearing mode models\cite{curie2022,curie2023microtearding} have been developed showing speed ups of almost two orders of magnitude over conventional models, and active learning pipelines have been shown to be able to create new training data efficiently \cite{zanisi2023efficient,activepipeline2022}. However, the physical simplifications and restricted parameter spaces used in these works, which help reduce the number of simulations required and the cost of those simulations, make them unsuitable for use in the predictive design of a spherical tokamak power plant. Consequently, new reduced-order models are required that are suitable for STs, both in terms of the parameter space explored and the fidelity of model used. 

Local linear gyrokinetic codes \cite{gkw_2009,gene,gs2} permit evaluation of the frequency, growth rates, normalised radial fluxes and the eigenfunctions of unstable modes for a given set of input parameters characterising the geometry and profiles at a particular flux surface. These are the building blocks of a reduced transport model \cite{Dudding_2022}. In this paper we aim to emulate these quantities using a database of gyrokinetic simulations of spherical tokamak power-plant relevant micro-tearing modes. To do this Gaussian Process Regression (GPR) is performed on an initially unexplored seven dimensional parameter space. A Gaussian process-based classifier is constructed which assists in the expansion of the training set by learning the stability manifold of the parameter space, allowing active learning techniques \cite{zanisi2023efficient,ChenKriging,Binois_2018,SeoGPR} to be employed which focus sample points on regions with high probability of instability to MTMs. The combined model is designed to investigate the parameter space as efficiently as possible while building an accurate regression based model harnessing the advantage of GPs that a prediction comes with an easily understandable confidence interval (CI) derived from the process variance.  

The remainder of the paper is outlined as follows: Section \ref{sec:gyro} briefly introduces the gyrokinetic model and the simulations used to generate training data; Section \ref{sec:inputs} defines the parameter space over which models will be built and Section \ref{sec:mtms} describes the modes of interest and the selection criteria used to identify them. An overview of Gaussian Processes is given in Section \ref{sec:datamodel} prior to a description of how they are used to build efficient models for MTMs in Section \ref{sec:active}. Validation of the resulting models is performed in Section \ref{sec:validation} and finally conclusions and ideas for future development are provided in Section \ref{sec:conclusions}. 

\section{Gyrokinetics}
\label{sec:gyro}

The gyrokinetic Vlasov-Maxwell system of equations are widely used to describe the turbulent component of the electromagnetic fields and particle distribution functions of the species present in a plasma. Here we will provide a short summary for understanding, but for a more comprehensive description of the model, see the following references \cite{Sugama_2000,BrizHahm2007,Krommes_2010}. To reduce the number of dimensions and the resolutions required to solve the equations, the following ordering assumptions are imposed which have been observed to be valid for turbulence in in the core of tokamak plasmas;
\begin{itemize}
    \item The Larmor radius $\rho_{a} = v_\perp/\Omega_{a} = v_{\perp}m_{a}/q_{a}B$ for species $a$, is small compared to the length scale over which the background kinetic profiles and magnetic equilibrium vary. The Larmor radius normalised to the system size (characterised by the minor radius $a$) $\rho_{*} = \frac{\rho_{a}}{a} \ll 1$ is thus considered a small parameter. 
    \item The turbulent fluctuations in a confined plasma are highly anisotropic due to the presence of a strong magnetic field, i.e the perpendicular wavelengths of interest are significantly smaller than those parallel to the magnetic field line, $k_{\perp} \gg k_{||}$.
    \item The fluctuations of the field perturbations against the background profiles are small. The electrostatic potential $\phi\approx\frac{T}{q_{a}}\rho_{*}$ and the electromagnetic potential $A_{||}\approx\frac{T}{q_{a}v_{th}}\rho_{*}$.
    \item The frequencies of interest of the turbulence and transport ($\omega$) are small compared to the ion cyclotron frequency, $\omega \ll \Omega_{a}=q_{a}B/m_{a}$.
\end{itemize}

The resulting 5-Dimensional gyrokinetic Vlasov equation has the form
\begin{equation}
    \frac{\partial F_{a}}{\partial t} + \frac{\partial\vec{X}}{\partial t}.\nabla F_{a}+\frac{\partial v_{||}}{\partial t} \frac{\partial F_{a}}{ \partial v_{||}}= 0,
    \end{equation}
where $F_{a}=F_{a}(\vec{x},v_{||},\mu)$ is the gyrocenter distribution function for species $a$ represented in full 5D phase space, $\vec{x}$ is the three-dimensional vector of coordinates of the guiding centre of a particle, $v_{||}$ is the velocity coordinate along the magnetic field line and $\mu= v_{\perp}^{2}/2B$ is the magnetic moment, a conserved quantity, where $v_{\perp}$ is the velocity in the perpendicular direction. A particle in the gyrokinetic framework is advected through configuration space by the plasma drifts caused by variations in the electric and magnetic fields along its trajectory, and accelerated along the magnetic field by the parallel electric field. 

The particle velocity can be written as:
\begin{equation}
    \frac{\partial\vec{X}}{\partial t} = \vec{v} =  v_{||}{\hat{b_{0}}}  + \vec{v}_{\nabla B} + \vec{v}_{\chi}
\end{equation}
where the terms on the right-hand side represent the parallel streaming, the drifts due to the inhomogenous magnetic field and the drifts due to the perturbed fields respectively. This equation is coupled to the gyrokinetic Poisson and Amp\`{e}re equations for closure.

Solution of the gyrokinetic Vlasov-Maxwell system enables an 
understanding of the stability of tokamak plasmas, and in the non-linear case, the saturated turbulent state and the resulting heat, particle and momentum fluxes. Nonlinear gyrokinetics represents one of the highest fidelity plasma turbulence modelling tools accessible on current supercomputers. It has been extensively validated against experiment \cite{white_2019,neiser_19,Hornsby_2017,Hornsby_2018} in a variety of regimes. Its large computational expense, however, makes it impractical for use in integrated modelling simulations over confinement time-scales, and as such quasilinear models such as TGLF\cite{Staebler_05,Staebler_07} and QuaLiKiz\cite{Bourdelle_2016} amongst others are used for this purpose. These models try to infer the nonlinear regime from the properties of the underlying linear modes. These are then used, along with a set of estimates of the saturation amplitudes of the mode based off of the underlying saturation mechanism, to estimate the nonlinear fluxes. However, the parameter regime of interest in this paper is outside the range of validity of such models, which were mostly developed for electrostatic turbulence in conventional aspect ratio tokamaks, and as such do not encapsulate the electromagnetic transport which is thought to be of primary importance in high $\beta$ spherical tokamaks.

\subsection{GS2}

GS2 is an implicit gyrokinetic code which enables solution of the gyrokinetic Vlasov-Maxwell system of equations using HPC resources and is used in the generation of training data for this study\cite{gs2}. It was specifically developed to study low-frequency turbulence in magnetized plasmas.  The code incorporates many physics effects  in the local limit including: kinetic electrons, electromagnetic effects (both $A_{||}$ and $B_{||}$ effects), collisions \cite{abel_08,barnes_09}, full general geometry and rotation.

It is typically used to assess the micro-stability of plasmas produced in the laboratory and to calculate key properties of the turbulence which results from instabilities. Linear micro-instability growth rates, frequencies and normalised fluxes can be calculated on a wavenumber-by-wavenumber basis using a flux tube geometry in the ballooning limit\cite{Beer_95}.  

GS2 has the option of being run as an initial value solver where the fastest growing mode for a given parameter set is found, or as an eigenvalue calculation where all the eigenvalue pairs for a parameter set can be found. The latter method has the advantage that all unstable modes can be found, including those that are subdominant. This is particularly useful for MTMs as they are in many cases subdominant to other microinstabilities, while still having a significant effect on heat transport. However, for ease of computation the former option is chosen, and usage of the eigenvalue solver is left for a future expansion of the model. The usage of an eigenvalue solver adds significant complexity to the classifier component of the model as will be described further in subsequent sections.

\section{Input parameter space definition}
\label{sec:inputs}

In this study a 7-dimensional parameter space is modelled comprising the variables considered to be those to which the MTMs are most sensitive. These parameters are:

\begin{itemize}
    \item The safety factor $q$: The ratio of toroidal to poloidal turns a field line of interest will wind around a flux surface.
    \item The magnetic shear $\hat{s}=\frac{r}{q}\frac{dq}{dr}$: Parameterises the radial variation of the safety factor, where $r$ is the radial coordinate. 
    \item The normalised radial gradient of the electron density $\frac{a}{L_{ne}}=-\frac{a}{n_e}\frac{\partial n_e}{\partial r}$. Where $a$ is the minor radius of the last closed flux surface.
    \item The radial gradient of the logarithmic electron temperature $\frac{a}{L_{Te}}=-\frac{a}{T_e}\frac{\partial T_e}{\partial r}$. 
    \item The ratio of the plasma pressure to the magnetic pressure $\beta = \frac{2\mu_{0}nk_{B}T}{B^{2}}$. Here the magnetic field is the toroidal magnetic field at the centre of the flux surface at the height of the magnetic axis. 
    \item The electron-ion collision frequency $\nu_{ei}$: The characteristic time scale at which electrons will scatter off ions.
    \item The normalised bi-normal wavevector $k_{y}$ (units of $\rho_{s}$).      
\end{itemize}

The ranges of the seven parameters are described in Table \ref{tab:variedparams} and represent plausible limits within which a future spherical tokamak power plant may be expected to operate. Other parameters, for instance those defining the flux surface shaping, are fixed at representative values based on STEP scenario SPR-008 \cite{stepbook} except for the ion temperature gradient, which is set to 0 to suppress ion temperature gradient (ITG) and trapped electron (TEM) modes which are not in the scope of this study and it is known to have a negligible effect on the MTM \cite{patel2023a}. The values of the fixed variables are given in Table \ref{tab:fixedparams}. The bi-normal mode wavevector is restricted to the range $0 \leq k_{y}\rho_{s} \leq 1$ for this initial model as this is the range of modes which dominate the heat flux spectra. An analysis of the mode stability for some of the flux surfaces that are considered here can be found in \cite{kennedy2023electromagnetic,giacomin2023electromagnetic}.

\begin{table}[!t]
\begin{tabular*}{\columnwidth}{@{\extracolsep\fill}llll@{\extracolsep\fill}}
\toprule
Variable & Name & Min.  & Max. \\
\midrule
$q$ & Safety factor  & 2   & 9  \\
$\hat{s}$ & Magnetic shear & 0   & 5  \\
$a/L_{ne}$ & Electron density gradient  & 0   & 10 \\
$a/L_{Te}$ & Electron temperature gradient  & 0.5 & 6  \\
$\beta$ & Ratio plasma/magnetic pressures                               & 0   & 0.3\\
$\nu_{ei}$ & Electron-Ion collision frequency ($c_{s}/a$)                            & 0   & 0.1\\
$k_{y}$ & Binormal mode wavelength (units of $1/\rho_{s}$)        & 0   & 1  \\
\end{tabular*}
\caption{Table showing the input parameters which are varied in this model with their maximum and minimum values.}%
\label{tab:variedparams}
\end{table}

\begin{table}[!t]
\begin{tabular*}{\columnwidth}{@{\extracolsep\fill}llll@{\extracolsep\fill}}
\toprule
Variable & Name & Value \\
\midrule
R  & Normalised major radius [m]      & 1.84  \\
a  & Minor radius [m] & 1.47  \\
$\kappa$ & Elongation          & 2.66  \\
$\frac{d\kappa}{d\rho}$ & Radial elongation gradient      & -0.25 \\
$\tau$   & Triangularity       & 0.34  \\
$\frac{d\tau}{d\rho}$ & Radial triangularity gradient     & 0.25  \\
$\Delta$   &  Shafranov shift                 & -0.44 \\
$\frac{T_{e}}{T_{i}}$  & Electron/ion temperature ratio      & 0.944 \\
$\beta'$  & Radial gradient of $\beta$   & -0.64 \\
$\theta_{0}$ & Ballooning angle & 0.0\\
\end{tabular*}
\caption{Table showing the parameters which are fixed in the model. Variations in these parameters will be considered in a future development.}
\label{tab:fixedparams}
\end{table}

\section{Micro-tearing modes and selection rules}
\label{sec:mtms}

\begin{figure}
    \centering
    \includegraphics[width=1.0\textwidth]{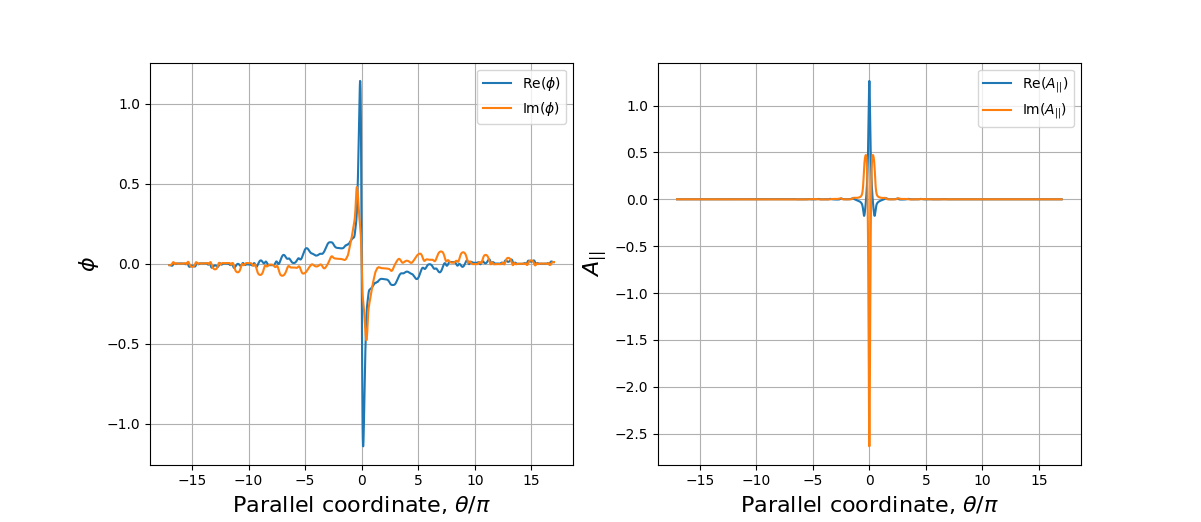}
    \includegraphics[width=1.0\textwidth]{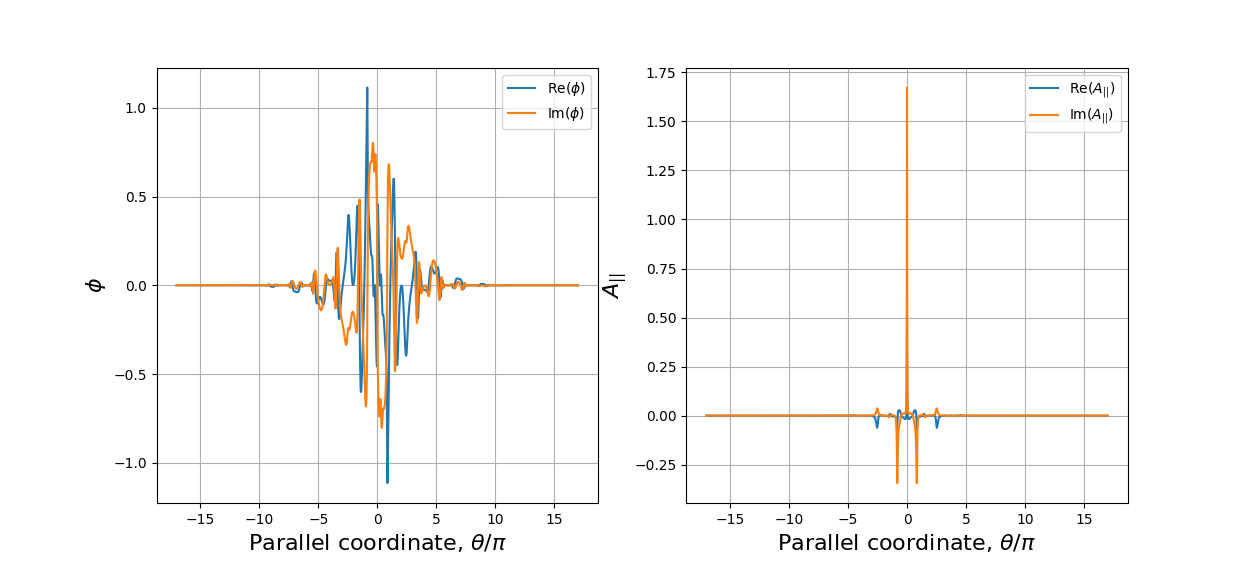}
    \caption{Example eigenfunctions of the electrostatic potential (left) and parallel magnetic potential (right) of an example of an MTM plotted along the parallel coordinate.  Real (blue) and imaginary (orange) parts are plotted.  The parameters  data point are, $k_{y}=0.384$, $\beta=0.204$, $a/L_{n}=1.072$, $a/L_{Te}=4.421$, $\nu_{ei}=0.075$, $q=2.92$, $\hat{s}=2.610$
    for the lower, $k_{y}=0.866$, $\beta=0.052$, $a/L_{n}=1.803$, $a/L_{Te}=4.473$, $\nu_{ei}=0.085$, $q=8.561$, $\hat{s}=1703$ }
    \label{fig:mtmexample}
\end{figure}
One of the characteristics of MTMs is their parity with respect to the coordinate along the magnetic field line. Specifically, the parallel magnetic vector potential ($A_{||}$) has an even parity around the outboard mid-plane ($\theta=0$), while the electrostatic potential ($\phi$) has an odd parity. Examples of this are seen in Figure \ref{fig:mtmexample}.  

\begin{figure} 
    \centering
    \includegraphics[width=1.0\textwidth]{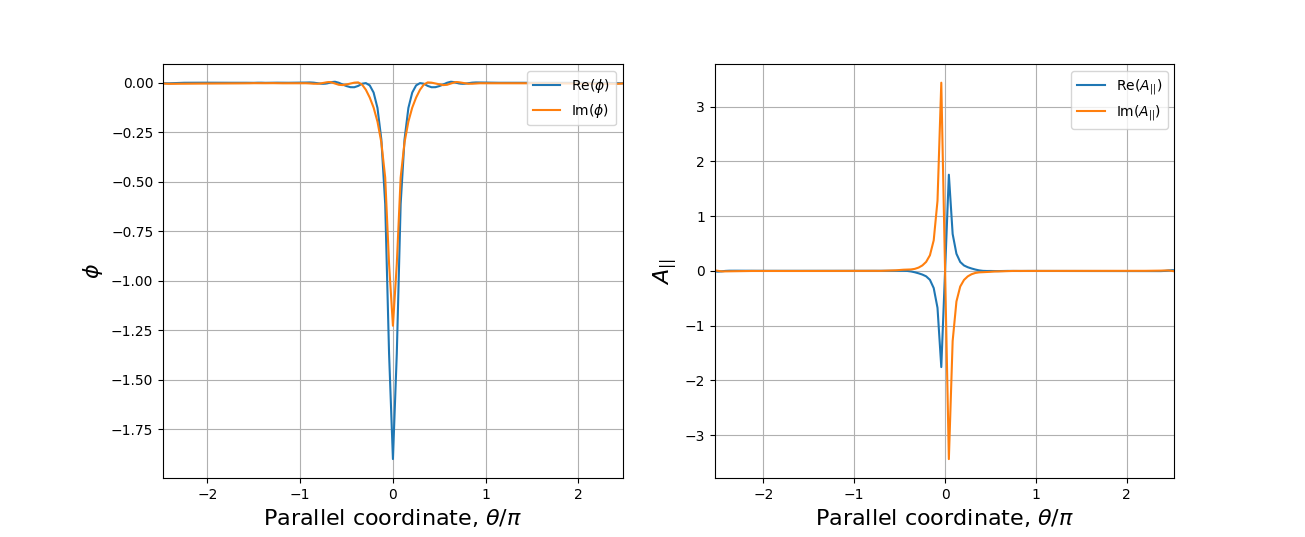}
    \includegraphics[width=1.0\textwidth]{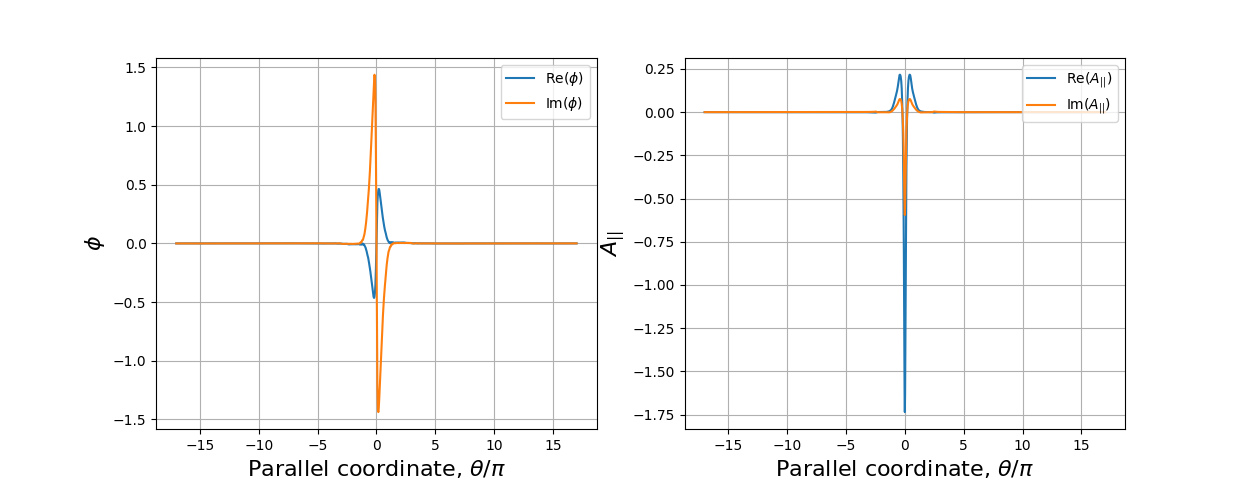}
    \caption{Example eigenfunctions of the electrostatic potential (left) and parallel magnetic potential (right) and of the kind of KBMs found plotted along the parallel coordinate.  Real (blue) and imaginary (orange) parts are plotted.  Top is an odd parity KBM (oKBM) (data point: $k_{y}=0.704$, $\beta=0.26$, $a/L_{n}=0.29$, $a/L_{Te}=4.23$, $\nu_{ei}=0.054$, $q=2.42$, $\hat{s}=2.43$), bottom is a KBM (datapoint: $k_{y}=0.053$, $\beta=0.166$, $a/L_{n}=4.468$, $a/L_{Te}=5.25$, $\nu_{ei}=0.07$, $q=8.56$, $\hat{s}=1.518$).  Setting $a/L_{Ti}=0$ filters out the upper modes, while the GS2 boundary parity filter, filters out the lower.}
    \label{fig:kbmexample}
\end{figure}

In order to ensure that GS2 predominantly finds micro-tearing modes, which in a lot of cases will be subdominant to ITGs, TEMs and KBMs, the desired parity can be enforced on the solution during the simulation.  This is done by setting the solution at negative $\theta$ to be minus the solution at positive $\theta$.  Even parity solutions can be forced in a similar way by setting the solution at negative $\theta$ to be identical to that at positive $\theta$.
This filters out modes with the parity seen in the lower panel of Fig. \ref{fig:kbmexample} and allows subdominant micro-tearing modes to be found even when running the code as an initial value solver. In general, for electromagnetic modes such as these, the magnitude of the magnetic perturbation $A_{||}$ is the same amplitude as, or larger than its electrostatic counterpart, $\phi$ ($|A_{||}| \geq |\phi| $) as seen in Fig. \ref{fig:mtmexample}.

These constraints do not guarantee that converged simulations will be MTMs, with large regions of parameter space having no unstable odd parity modes and odd parity kinetic ballooning type modes (KBMs) also being present, such as those seen in the upper panel of Fig. \ref{fig:kbmexample}. To filter these out of our training data, additional selection criteria are set which are developed from knowledge of the physics of the expected modes. These are applied to the set of converged runs to produce a training set identified as MTMs. MTMs perturb field lines so that they do not return to the equilibrium flux surface. This effect can be quantified by the \textit{tearing parameter}:

\begin{equation}
C_{tear} = \frac{\left|\int A_{||}d\theta\right|}{\int\left| A_{||}\right|d\theta}
\end{equation}

where $\int d\theta$ corresponds to the integral along the field line, which is parameterised by the field following co-ordinate, $\theta$. This parameter is calculated for each simulation and those with values less than 0.15 are rejected. 

Micro-tearing modes have frequencies that are in the electron diamagnetic direction \cite{hazeltine1975}. By definition these have a negative sign in GS2. This allows a second selection rule; the calculated frequency of the mode must be within a window whose bounds are 50\% above and below the value predicted by the analytical lowest order expression as derived by Catto {\it et al}\cite{catto1981}:

\begin{equation}
\omega_{0} = -(k_{y}\rho_{s})\left(\frac{a}{L_{ne}} + 0.5\frac{a}{L_{Te}} \right)
\label{eq:mtmfreq}
\end{equation}

\begin{figure}
    \centering
    \includegraphics[width=0.8\textwidth]{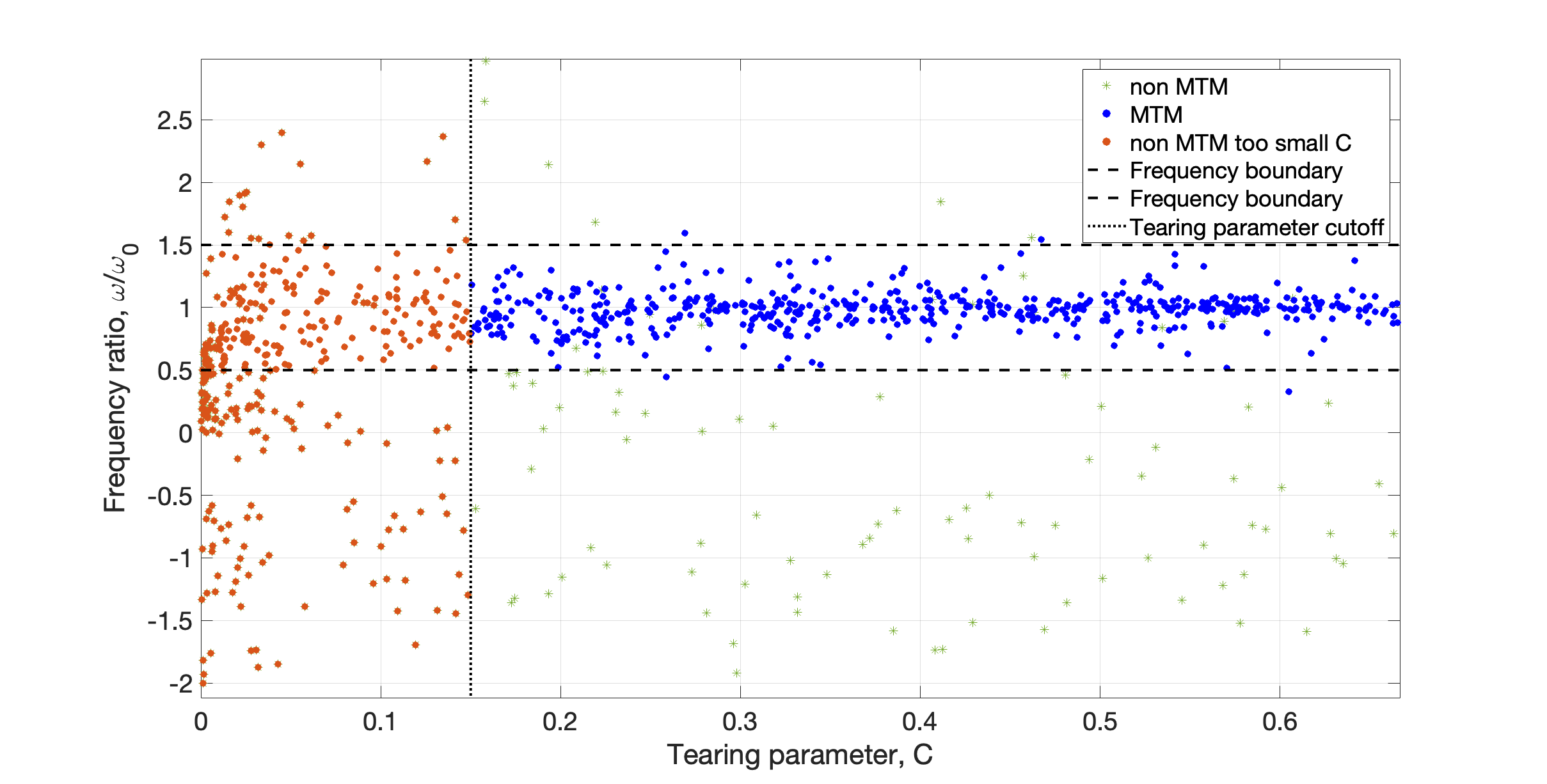}
    \caption{Distribution of points in tearing parameter and frequency (normalised to the analytical expression, $\omega_{0}$, Eq. \ref{eq:mtmfreq}) space for a subset of the MTM database. Blue points are accepted into the training database, while the green crosses and red points are not for being outside the acceptance criteria as denoted by the horizontal (frequency) and vertical (tearing parameter) dark lines.}
    \label{fig:tearingdistribution}
\end{figure}

Figure \ref{fig:tearingdistribution} shows the distribution of a subset of the converged simulations in normalised mode frequency and tearing parameter space. The horizontal lines represent the lower and upper bounds of the normalised frequency within which an MTM is expected to sit, and the vertical dashed line represents the lower cut-off to the tearing parameter. A final visual inspection of the structure of the eigenfunctions removes points which are unconverged or lack sufficient length along the field line to fully resolve the mode, this rejects the non-MTM (blue) points within the acceptance limits in Fig. \ref{fig:tearingdistribution}. These physics-based selection rules are used to set a logical parameter indicating whether a converged instability is considered to be a MTM or not. It is upon this parameter that the classifier component of the model is trained.  

\subsection{GS2 simulation setup}

The MTMs will be examined using local linear gyrokinetics on a single core flux surface with $\rho = r/a = 0.67$, where $a$ is the radius of the last closed flux surface. A Miller parameterisation \cite{miller1998} was used to model the equilibrium, with the parameters outlined in Table \ref{tab:fixedparams}. Three kinetic species are modelled: deuterium, tritium and electrons ($n_{e} = n_{D} + n_{T}$, deuterium and tritium in a 50/50 mix). The simulations are run for a single bi-normal mode, $k_{y}$. We neglect compressional magnetic field ($B_{||}$) effects in order to reduce the computational complexity of the model and suppress some types of kinetic ballooning mode which are destabilised by compressional magnetic field perturbations. Micro-tearing modes are essentially unaffected by their inclusion. 


\begin{table}[!t]
\begin{tabular*}{\columnwidth}{@{\extracolsep\fill}llll@{\extracolsep\fill}}
\toprule
Variable & Description  & Value \\
\midrule
$n_{\epsilon}$ & Number of energy grid points.   & 16  \\
$n_{||}$ & Number of un-trapped pitch angle grid points \\  
& in each direction along the
field line.   & 16  \\
$n_{trapped}$ & Number of grid points in the trapped region \\
& of velocity space. & 65 \\
$n_{\theta}$ & Sets the number of parallel grid points along \\
& the magnetic field line per poloidal turn. & 128 \\
$n_{period}$ & Defines the domain of the simulation by setting the  \\ & number of times the flux tube winds around the \\
& device poloidally. & 9
\end{tabular*}
\caption{Table showing the resolutions used the parallel direction and velocity grids when generating the training data.}
\label{tab:resparams}
\end{table}

MTMs require significant numerical resolution due to being particularly elongated along the magnetic field line in ballooning space, equivalent to having a very fine radial structure. A single set of numerical inputs is used in this work for tool-chain simplicity. These are summarised in Table \ref{tab:resparams}. $n_{period}$ defines the domain of the simulation by setting the number of times the flux tube
winds around the device poloidally. The number of $2\pi$ segments is given by 2$n_{period} - 1$. GS2 uses the pitch angle and particle energy as velocity space coordinates.  $n_{||}$ determines the number of un-trapped pitch angle grid points. The number of trapped pitch angle points is defined by the resolution of the $\theta$ grid as $n_{\theta}/2 + 1$.

The run time of the simulation is approximately linear in the value of $n_{period}$ so a value of 9 was chosen for this work as a compromise between accuracy and required computational resources, however there will be a systematic uncertainty resulting from the choice of this parameter, especially in parts of parameter space where the mode is particularly elongated in the ballooning angle. The simulations where this effect is particularly acute are removed from the training set to prevent such systematic errors from entering the regression model. The resolutions used here are influenced by, and consistent with, the resolutions used in previously published studies of the linear structure of unstable modes in spherical tokamaks \cite{patel2021}. It is noted here that this resolution is approximately an order of magnitude higher than for the KBMs or trapped electron modes (TEM) that would also be found in this parameter space where 2/3 poloidal turns is usually sufficient, and the modes tend to be more strongly growing.

\subsection{Model outputs}

The outputs of the model are the properties of the micro-tearing modes that are important for building a quasi-linear heat transport model. Namely the growth rates, frequencies and mode contribution to the radial electron heat flux caused by the interaction of the turbulent field and the turbulent pressure fluctuations. These are outlined in Table \ref{tab:modeloutputs} along with their units.

The heat flux is defined as, 
$Q_{em,e} = \langle\int d^{3}{v} \frac{mv^{2}}{2}{\tilde{v_{\delta B}}}\cdot\nabla{r}f\rangle$, normalised to $\langle|A_{||}|^{2}\rangle$, where $\tilde{v}_{\delta A_{||}} = \hat{b}\times\nabla{(v_{||}\tilde{A_{||}})}/B $ is the flutter velocity, the parallel velocity along the perturbed field lines, and the curly brackets denote a flux surface average (Where for quantity $F$, $\langle F\rangle= \int F dl$). $\hat{b}$ denotes a unit vector along the equilibrium magnetic field and $B$ is the magnetic field strength. This is the predominant radial heat transport driven by MTMs. 

In addition to GS2 our tool-chain incorporates Pyrokinetics \cite{pyrokinetics} which is a python library that aims to standardise gyrokinetic analysis by providing a single interface for reading and writing input and output files from different gyrokinetic codes, normalising to a common standard. This enables future interoperability with other gyrokinetic codes. All variables in this paper utilise the standard normalisations used by Pyrokinetics \cite{pyrokinetics}.

\begin{table}[!t]
\begin{tabular*}{\columnwidth}{@{\extracolsep\fill}llll@{\extracolsep\fill}}
\toprule
Variable & Name & Units \\
\midrule
$\gamma$  & Mode growth rate      & $c_{s}/a$  \\
$\omega$  & Mode frequency & $c_{s}/a$  \\
$Q_{em,e}$ & Radial electromagnetic electron heat flux          & $\rho_{*}^{2}n_{ref}c_{s}T_{e}$  \\
\end{tabular*}
\caption{Model outputs.  $a$ is the minor radius of the Tokamak, while  $c_{s}=\sqrt{T_{i,ref}/m_{e}}$ is the ion sound speed. The electron heat flux is related to the heat diffusivity and temperature gradient by $Q_{e} = -n_{e}\chi_{e}\nabla T_{e}$.}
\label{tab:modeloutputs}
\end{table}
 
\section{Gaussian processes}
\label{sec:datamodel}

Our data-driven surrogate model of the linear properties of MTMs is built on Gaussian process (GP) regression.  An introduction can be found here \cite{rasmussen_gaussian_2006}. GP regression is a Bayesian approach that has received significant interest in recent years and has been used in the modelling of a diverse range of physical systems. A Gaussian Process defines a prior over a function space which can be conditioned on observations of the function to be emulated. Once trained it provides a probability distribution over the function value at any unobserved location, allowing prediction with a quantified uncertainty.

Gaussian Process regression is used as opposed to neural networks for the model as they are readily interpretable and, importantly when doing uncertainty quantification, they handle uncertainty and noise in a rigorous fashion, two properties that are not always applicable to neural networks. Due to the expense of running a gyro-kinetic code, even linearly, the dataset is expected to be small, and as such Gaussian processes have an advantage due to their data efficiency.

\begin{figure}
    \centering
    \includegraphics[width=1.0\textwidth]{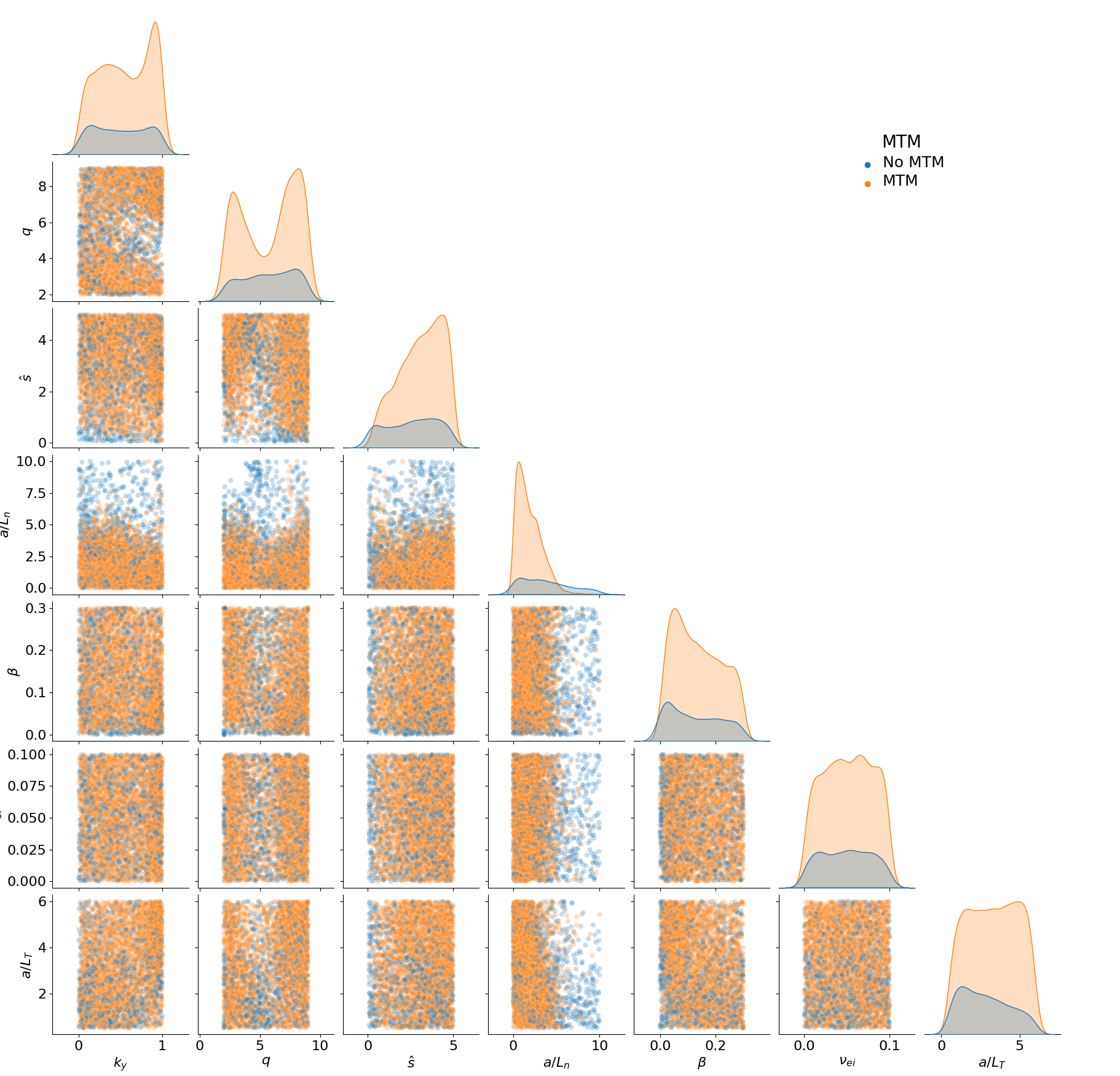}
    \caption{The distribution of simulations over the seven dimensional domain. The orange points represent simulations where an unstable micro-tearing mode is found while blue points are those found to be stable. It is evident that the electron density gradient is highly stabilising, as there are few unstable points seen when $a/L_{n}>4$, however the stability region is otherwise quite complex, with significant structure over the parameter space. Plots on the diagonal show the marginal distributions of stable and unstable points in each parameter.}
    \label{fig:databasescatter}
\end{figure}

\subsection{Gaussian Process Regression}

A Gaussian Process is a stochastic process (a collection of random variables) such that every finite subset has a joint probability distribution which is Gaussian. In the context of function emulation one can consider the value of the function $f$ at each location in the input space to be given by a random variable, such that the function as a whole is comprised of an infinite collection of random variables which obey the above property. This effectively yields a probability distribution for the function which can then be conditioned on observations. 

The joint probability distribution for the function at a finite set of input locations is given by a multivariate Gaussian as follows: 
\begin{equation}
f \sim \mathcal{N}(\mu(x),\,\Sigma(x,{x}'))
\end{equation}
for any two points ${x}$, ${x}'$ in a set $X$, $\mathcal{N}$ denotes a Gaussian distribution where $\mu$ is the \textit{mean function} and $\Sigma$ is the \textit{covariance function}, which is a positive semi-definite function which maps pairs of locations in parameter space to the covariance between their function values.  

If there is prior knowledge of the overall structure of the function to be emulated, then this can be expressed via the choice of parameterisation of the mean function, but in many cases, this is simply set to zero. In this work, we use a constant mean function, unless otherwise stated.

The posterior mean and variance are given by:
\begin{eqnarray}
\mu(x_{\star}) = \kappa({x},{x}_{\star})^{T}(\kappa({x},{x})+\sigma^{2}I)^{-1}f(x)\\
\Sigma({x},{x}_{\star}) = \kappa({x}_{\star},{x}_{\star}) -  \kappa({x},{x}_{\star})^{T}(\kappa({x},{x})+\sigma^{2}I)^{-1}\kappa({x},{x}_{\star})
\end{eqnarray}
where $x_{\star}$ denotes a vector of parameter space locations to be predicted. Where $\sigma^{2}$ is the process variance and $\kappa$ is a kernel function modelling the correlation between data-points.  $\kappa({x},{x})$ denotes a $n\times n$ matrix of the covariances between points in a training set of size $n$.  Similarly, $\kappa({x},{x_{\star}})$ is a matrix of $n\times n_{\star}$ size containing the covariances between all pairs of training and test points of number $n_{\star}$.

In relation to our model, the mean $\mu$ corresponds to the growth rate, mode frequency and fluxes for a specific input parameter set $x_{\star}$ (the seven input parameters as outlined in Table \ref{tab:variedparams}).  The variance $\Sigma$ represents the uncertainty in those predicted mean values. 

\subsubsection{Kernels}

The covariance function is expressed as the product of $\sigma^{2}$ and $\kappa$:

\begin{equation}
\Sigma({x},{x}') = \sigma^{2} \kappa({x},{x}')
\end{equation}

Many choices are available for the kernel function, but two common choices used when one has no prior knowledge of the covariance structure are the radial basis function (RBF) kernel: 

\begin{equation}
\kappa({x},{x}') = \Pi_{i=1}^{p}\exp(-\frac{({x}_{i}-{x}_{i}')^{2}}{2l_{i}^{2}})
\end{equation}
 
\noindent and the Mat\'{e}rn family of kernels parameterised by $\nu$: 

\begin{equation}
\kappa({x},{x}') = \Pi_{i=1}^{p} \frac{1}{2^{\nu-1} \Gamma_{\nu}}  \left(\frac {2\nu^{1/2} |{x}_{i}-{x}'_{i}| } { l_{i} } \right)^{\nu} J_{\nu} \left(\frac {2\nu^{1/2} |{x}_{i}-{x}'_{i}| } { l_{i} } \right)
\end{equation}

where $\Gamma_{\nu}$ is the Gamma function for $\nu$, $J_{\nu}$ is a modified Bessel function of order $\nu > 0$ and $p$ is the number of input dimensions, as outlined in Table \ref{tab:variedparams}. In both cases, each dimension $i$ has an associated correlation scale length given by $l_{i}$. Points in parameter space much closer than these characteristic distances can be expected to be strongly correlated. 

The squared exponential in the RBF kernel is infinitely differentiable, which can lead to unrealistically smooth function realisations. The Matérn kernel is only $\lfloor \nu - 1 \rfloor$ times differentiable and thus allows more flexible modelling of the degree of smoothness of the underlying function. Note that the Matérn kernel can be simplified for $\nu = p + \frac{1}{2}; p \in \mathbb N_0$, yielding particularly `nice' solutions for $\nu = [1/2, 3/2, 5/2]$. When $\nu \geq 7/2$, the kernel is so smooth that it is hardly distinguishable from the RBF kernel \cite{rasmussen_gaussian_2006}. The Matérn 5/2 kernel is a common choice for emulation and the default kernel used in the experiments below. 

\subsubsection{Fitting to the data}
Before seeing any data, the structure outlined above constitutes the GP \textit{prior}, and it specifies a certain set of functions according to the mean function and the kernel. This prior over functions $f$ for given \textit{hyperparameters} $\theta$ can be written compactly as
\begin{equation}
    p(f | \Theta)
\end{equation}
where the model hyperparameters $\Theta$ are the covariance length scales of the kernel (denoted $l_{i}$ above) and the noise level. The aim is to identify the posterior distribution after conditioning the prior on inputs $X$ and outputs $y$ using Bayes' Theorem:
\begin{equation}
    p(f | y, X, \Theta) = \frac{p(f | \Theta) \: p(y | X, f)}{p(y | X, \Theta)}
\end{equation}
where the \textit{marginal likelihood} is
\begin{equation}\label{eq:marginal_likelihood}
    p(y | X, \Theta) = \int p(f | \Theta) \: p(y | X, f) \: df
\end{equation}
This is relatively straightforward for fixed hyperparameters, but these are rarely known {\it a priori} and hence a similar approach must be taken at the next higher level of this hierarchical model to identify the posterior distribution over model hyperparameters:
\begin{equation}\label{eq:hyper-posterior}
    p(\Theta | y, X) = \frac{p(\Theta) \: p(y | X, \Theta)}{p(y | X)}
\end{equation}
with marginal likelihood
\begin{equation}
    p(y | X) = \int p(\Theta) \: p(y | X, \Theta) \: d\Theta
\end{equation}
inferring the hyperparameters from the data by marginalising over model hyperparameters $\Theta$ is challenging since this integral rarely has an analytical solution. This can be alleviated through approximations, such as Variational Inference (VI), or by sampling using Markov chain Monte Carlo (MCMC). In practice, Type-II Maximum Likelihood Estimation (MLE) is often employed. Instead of evaluating or approximating the full posterior of hyperparameters (Eq.~\ref{eq:hyper-posterior}), the marginal likelihood (Eq.~\ref{eq:marginal_likelihood}) is maximised with respect to the hyperparameters $\Theta$. Thus, the hyperparameters will be fixed at the maximum likelihood values. Please refer to \cite{rasmussen_gaussian_2006} for more details. Due to the relatively large datasets used in this study, Type-II MLE will be employed for regression throughout. Irrespective of the method employed to approximate the model hyperparameters, fitting a GP requires inverting an $n\times n$ dense matrix for $n$ data points. Hence, GP regression scales extremely poorly with increasing numbers of datapoints with a compute cost of $\mathcal{O}(n^3)$. In this work Blackbox Matrix-Matrix multiplication (BBMM) is leveraged to dramatically reduce this cost, reducing it to $\mathcal{O}(n^2)$, as described in \cite{gardner_gpytorch_2018}.

\subsubsection{Gaussian Process Classification}
The classification problem consists of determining whether or not a certain input would result in an unstable MTM. This is a \textit{binary} classification problem, which can be handled using a GP with a Bernoulli distribution as the likelihood and a \textit{probit} link function, i.e. the Standard Normal Cumulative Density Function (CDF) $\Phi(\cdot)$. This probit Gaussian Process classification model can be written compactly as
\begin{equation}
    p(\text{MTM}=1|x) = \Phi(f(x))
\end{equation}
where $f(x)$ is a sample drawn from a Gaussian Process, commonly known as a \textit{latent} function or \textit{nuissance} function. This results in a stochastic classification model with a smoothly varying decision boundary. The procedure for fitting the latent function is similar to the approach outlined above for GP regression, however, the posterior for this model is analytically intractable and must be approximated either through sampling methods, such as MCMC, or VI. While the posterior recovered by MCMC is asymptotically exact, it can be prohibitively computationally expensive to perform this sampling, particularly for large datasets. Hence, VI is often employed for approximate inference. Here, the posterior is approximated by minimising the Kullback-Leibler (KL) divergence between the posterior and a computationally tractable distribution, usually referred to as the \textit{variational} distribution \cite{hensman2015scalable}. Please refer to \cite{rasmussen_gaussian_2006} for more details.

After conditioning on training data, the $\Phi$-transformed posterior mean of the latent function can be interpreted as the probability of a test point $x_\star$ being an MTM, i.e. $p(\text{MTM}=1|x_\star) = \Phi(\mu(x_\star))$ and conversely $p(\text{MTM}=0|x_\star) = 1 - \Phi(\mu(x_\star))$. This classifier will enable efficient expansion of the training set by predicting the regions of parameter space where an MTM may be found and thus allowing a targeted use of compute resources. The classifiers were evaluated using standard metrics, namely accuracy, precision, recall and F-score (see e.g. \cite{taha_metrics_2015}).

\section{Active learning workflow} 
\label{sec:active}

\subsection{Parameter space exploration}

\begin{figure}[h!]
    \centering
    \includegraphics[width=0.8\textwidth]{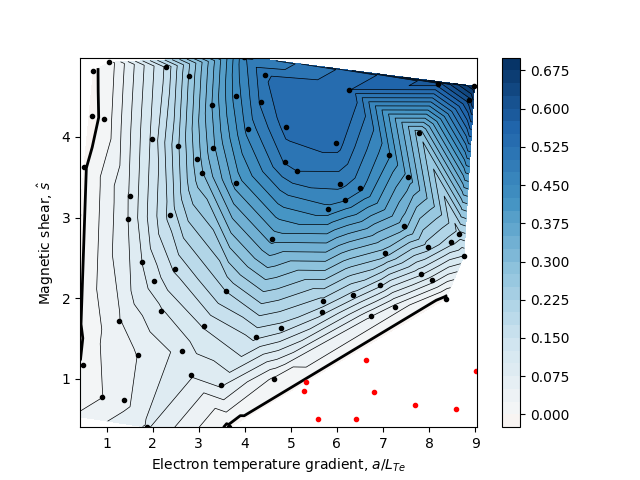}
    \caption{The MTM growth rate calculated as a function of the electron temperature gradient ($a/L_{Te}$) and the magnetic shear ($\hat{s}$) using a 100 point Latin hypercube sample with all other parameters kept constant. Black points are simulations where an MTM is found whereas red points are observed non-MTMs and some MTMs with a negative growth rate were found. The dark black lines are the contours of zero growth rate, the marginal stability manifold. Distinct areas of stability and instability and evident.}
    \label{fig:2dplane}
\end{figure}

The microtearing mode is stabilised by some of the parameters that are investigated here, and as such there are large volumes of parameter space where GS2 may not return a positive result as is evident in the locations of the simulation points seen in Fig.~\ref{fig:databasescatter}. To demonstrate this, Fig.~\ref{fig:2dplane} shows a contour plot of the calculated growth rate across a 2-dimensional parameter space of magnetic shear $\hat{s}$ and electron temperature gradient $a/L_{Te}$ calculated using a Latin hypercube with 100 samples. Stability boundaries where the growth rate of the MTMs tends to zero are evident in the bottom right-hand corner and additionally at low electron temperature gradient. This submanifold characterising marginal stability is of great importance for determining the parameter values beyond which turbulence can form.

As GS2 is being used as an initial value code, in cases where the simulation is run in areas of stability it will run to a pre-determined maximum time without finding an instability, thus using up significant computational resources without producing useful data upon which to train the regression model. To mitigate this, a classifier is trained to predict the probability that a given point in parameter space will be unstable to MTMs, effectively learning the marginal stability manifold and allowing targeted exploration of the parameter space. To build such a classifier an initial data set is required.   

\subsubsection{Initial data-set generation} \label{sec:initial_data}

An initial data set was produced by generating a 300 point maxi-min Latin Hypercube design over the parameter space. A Latin Hypercube is a space filling quasi-Monte-Carlo sampling scheme whereby one partitions each parameter range into N divisions and samples N points such that each such partition only contains one sampled point for every parameter. A maxi-min design attempts to do this in a way that maximises the minimum separation between points in order to generate a point set with good coverage and minimal clustering.

Out of the 300 points sampled, 293 converged and of these, 99 passed the selection criteria for MTMs discussed in Section \ref{sec:mtms} and in total cost approximately 9000 CPUhrs. The classifier was trained on both the samples identified as MTMs and those identified as stable (no MTM) while the regression models were trained solely on the MTMs. The training of the classifier takes approximately one hour on a single CPU.  The low fraction ($\sim30\%$) of MTMs produced by the Latin hypercube, a uniform sampling approach, and the high associated computational overhead, highlights the potential benefit of using an active learning approach. 

\subsection{Additional data points for GP regression}
\label{sec:activelearning}

To improve the predictions of the GP regression, more samples were drawn iteratively from areas of the parameter space with high posterior probability of finding an MTM as determined by the classifier. The algorithm is summarized as follows:

\begin{itemize}
    \item Draw $n = 2^k$ low discrepancy candidate data points from the input space using Sobol sequences \cite{sobol_distribution_1967} to yield an initial sample $X = \{x_i\}_{i=1}^n$.
    \item Evaluate the posterior predictive probability of being an MTM for each candidate data point using the Bernoulli GP classifier $p(\mathrm{MTM}=1|x_i) = \Phi(\mu(x_i))$, where $\Phi(\cdot)$ is the Standard Normal CDF, and $\mu(x_i)$ is the predictive mean of the latent function for data point $x_i$.
    \item Accept/reject each candidate data point using a rejection sampler on the posterior predictive probability of being an MTM, i.e.
    \begin{itemize}
        \item Draw a random number from a uniform distribution $\rho_i \sim \mathcal U(0,1)$.
        \item If $\rho_i < p(\mathrm{MTM}=1|x_i)$, accept the candidate data point $x_i$.
        \item Otherwise, reject the candidate data point $x_i$.
    \end{itemize}
    \item This algorithm yields a final sample $X^\star$ of accepted samples, where the samples are now distributed according to the posterior predictive distribution of MTMs, $p(\mathrm{MTM}=1|x_i)$.
    \item Retrain the classifier and regressors using the expanded data set. 
    \item Iterate until the desired performance is achieved or the available compute resources are expended.
\end{itemize}

\begin{figure}[htbp]
    \centering
    \includegraphics[width=1.0\textwidth]{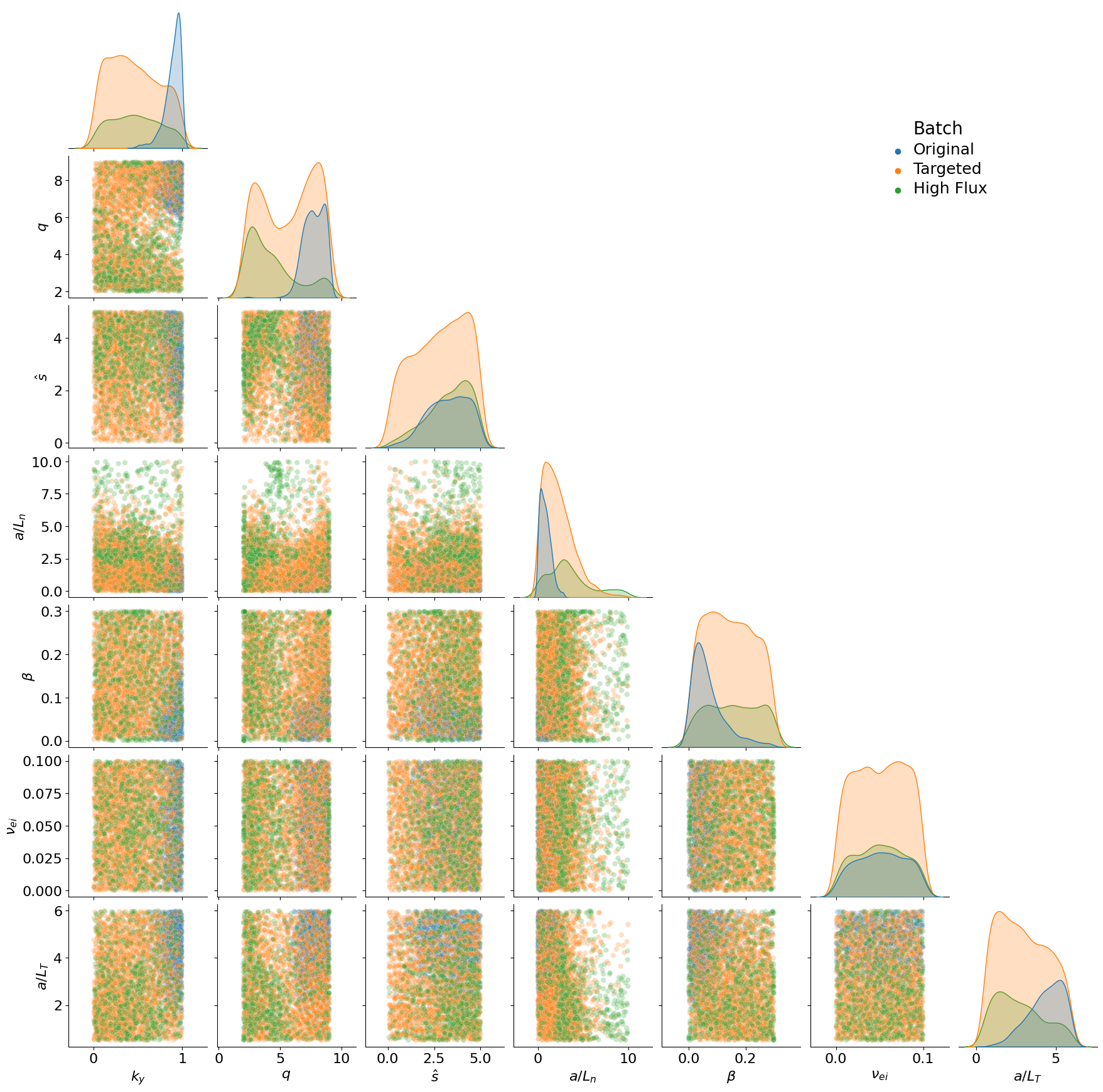}
    \caption{The complete dataset with each data point coloured by its respective batch. The original dataset (blue) is the data produced using a 300 point Latin hypercube design as outlined in Section \ref{sec:initial_data}. The targeted dataset (orange) encompasses both batches of samples directed towards finding an MTM. The high flux datasets (green and red) were from areas of high predicted flux, $Q_{em,e} > 0.2$ and $Q_{em,e} > 0.5$ respectively and are combined here denoted as High Flux.}
    \label{fig:pairplot_batches}
\end{figure}

Two batches of new data were generated using this procedure. The first batch consisted of $1362$ accepted samples, of which $971$ were subsequently identified as MTMs, corresponding to a hit rate of $71\%$. The second batch consisted of $1317$ accepted samples, of which $983$ were subsequently identified as MTMs, corresponding to a hit rate of $75\%$. This is a significant increase compared to the initial parameter space sampling, representing a substantial saving in computational time. A summary of the batch statistics can be found in Table~\ref{tab:batchstats} and the locations of the simulation points in parameter space is showing in figure \ref{fig:pairplot_batches}.

\begin{table}[!t]
\centering
\footnotesize
\begin{adjustbox}{angle=90}
\begin{tabular*}{\columnwidth}{lrrrrr}
\toprule
$N^{o}$ & Name & Description & Tot. num.  & Number of  & Hit rate  \\
& & & simulations & MTMs & (\%) \\
\midrule
0 & Initial LHD & Original LHD distributed  & 293 & 99 & 33 \\
&& parameter scan &&\\
1 & Original & First expansion of dataset  & 1000 & 513 & 51.3  \\
& & using rudimentary classification &&\\
2 & Targeted1  & General classifier directed  & 1361 & 971 & 71.2  \\
& & database expansion & & & \\
3 & Targeted2  & General classifier directed  & 1316      & 983 & 74.6  \\
 & & database expansion & & & \\
4 & HighFlux1 & Dataset targeted to  & 300         & 237  & 79  \\
& & high flux region ($Q_{em,e} > 0.2$) & & & \\
5 & HighFlux2 & Dataset targeted to  & 400 & 349 & 87 \\
& & high flux region ($Q_{em,e} > 0.5$) & & & \\
\bottomrule
\end{tabular*}
\end{adjustbox}
\caption{Table outlining the number of simulations, the number of simulations found with an unstable MTM and the percentage of the batch that gives an unstable MTM for the training set. Evident is the increase in hit rate as more data is added to the database and the classifier fidelity improves. }
\label{tab:batchstats}
\end{table}

The model trained on this expanded data set was observed to lack accuracy at high heat fluxes and so further iterations targeting high flux points were performed. To do this the existing data were labelled according to whether they were below or above a specified heat flux threshold and a Bernoulli GP classifier was trained on these labels. Sample points with a high predicted heat flux value were then proposed using the rejection sampling algorithm outlined above, highlighting the flexibility of the classifier driven sampling approach. Using this approach two further iterations targeting $Q_{em,e} > 0.2$ and $Q_{em,e} > 0.5$ were performed.

\section{Validation}
\label{sec:validation}

\subsubsection{Diagnostics}
The predictive power of each model was evaluated using the Mean Standardised Log Loss (MSLL) metric \cite{rasmussen_gaussian_2006}. This metric computes the average negative log-probability of the testing data with respect to their predictive posterior densities, standardised by the loss of the trivial model $y_\star \sim \mathbb{N}(\mu_\mathcal{D}, \sigma_\mathcal{D}^2)$:
\begin{align}
    - \log p(y_\star|\mathcal{D}, x_\star) &= \frac{1}{2} \log(2 \pi \sigma_\star^2) + \frac{(y_\star - \bar{f}(x_\star))^2}{2\sigma_\star^2} \\
    \mathrm{MSLL(y_\star)} &= - \log p(y_\star|\mathcal{D}, x_\star) + \log \mathbb{N}(y_\star; \mu_\mathcal{D}, \sigma_\mathcal{D}^2)
\end{align}
where $(x_\star, y_\star)$ is a test point, $\mathcal{D} = \{(x_i, y_i)\}_{i=1}^n$ is the training dataset, $\sigma_\star^2$ is the predictive variance and $\mu_\mathcal{D}$ and $\sigma_\mathcal{D}^2$ are the mean and variance of the training data, respectively.

Additionally, the calibration error of each model was evaluated using the Root Mean Squared Calibration Error (CE) \cite{kuleshov_accurate_2018,chung2021uncertainty, tran2020methods}, which is a frequentist method that measures the difference between the expected and the observed cumulative distributions of model predictions:

\begin{align}
    \nu(q) &= \frac{1}{N}\sum_{i=1}^N \mathbb{I} (y_i \leq \mathcal{Q}_{x_i}(q))\quad q \in [0,1]\\
    \mathrm{CE} &= \sqrt{\frac{1}{M}\sum_{i=1}^M (q_i - \nu_i)^2}
\end{align}
where $\mathbb{I}(\cdot)$ is the indicator function which returns 1 if the argument is true and 0 otherwise, $\mathcal{Q}_{x_i}(q)$ is the quantile function of the posterior prediction of $x_i$ evaluated at the quantile $q \in [0,1]$, so that $\nu(q)$ is the average number of datapoints within a given quantile $q$. For a perfectly calibrated model $\nu(q) \to q \: \forall \: q \in [0,1]$ as $N \to \infty$. Then the CE can be computed from $M$ samples of $q$ and $v$, $\{(q_i, \nu_i)\}_{i=1}^M$. The remaining regression metrics are explained in \ref{sec:glossary}.

\subsection{Classification}
The Bernoulli GP (BGP) classifier was benchmarked against standard reference classifiers, namely Logistic Regression (LR) (see e.g. \cite{bishop_pattern_2006,hastie_elements_2009}) and a Gradient-Boosting Classifier (GB) \cite{friedman_greedy_2001,friedman_2002,hastie_elements_2009}. The BGP was tested using three different covariance kernels, namely Matérn 1/2, 3/2 and 5/2 (M12, M23 and M52). Table \ref{tab:classifiation_benchmark} shows the mean and standard deviation of the accuracy, precision, recall and F1 score for each classifier. For definitions of these terms the reader is directed to \ref{sec:glossary}. Here, the mean and standard deviation were computed from a 5-fold cross-validation. Note that the Bernoulli GP Classifiers have slightly lower accuracy and precision, and slightly higher recall than the reference classifiers. 

\begin{table}[htbp]
\centering
\footnotesize
\begin{tabular}{lrrrr}
\toprule
Classifier & Accuracy & Precision & Recall & F1 \\
\midrule
LR        & 0.827 ± 0.027 & 0.813 ± 0.019 & 0.907 ± 0.036 & 0.857 ± 0.024 \\
GB        & 0.879 ± 0.036 & 0.868 ± 0.046 & 0.935 ± 0.024 & 0.900 ± 0.026 \\
BGP (M12) & 0.875 ± 0.016 & 0.847 ± 0.024 & 0.954 ± 0.008 & 0.898 ± 0.014 \\
BGP (M32) & 0.888 ± 0.026 & 0.865 ± 0.029 & 0.955 ± 0.018 & 0.907 ± 0.020 \\
BGP (M52) & 0.890 ± 0.017 & 0.864 ± 0.032 & 0.961 ± 0.013 & 0.909 ± 0.015 \\
\bottomrule
\end{tabular}
\caption{Mean and standard deviation of the accuracy, precision, recall and F1 score for the Logistic Regression (LR), Gradient Boosting Classifier (GB) and Bernoulli GP Classifier (BGP).}
\label{tab:classifiation_benchmark}
\end{table}

\begin{figure*}
    \centering
    \includegraphics[width=1.0\textwidth]{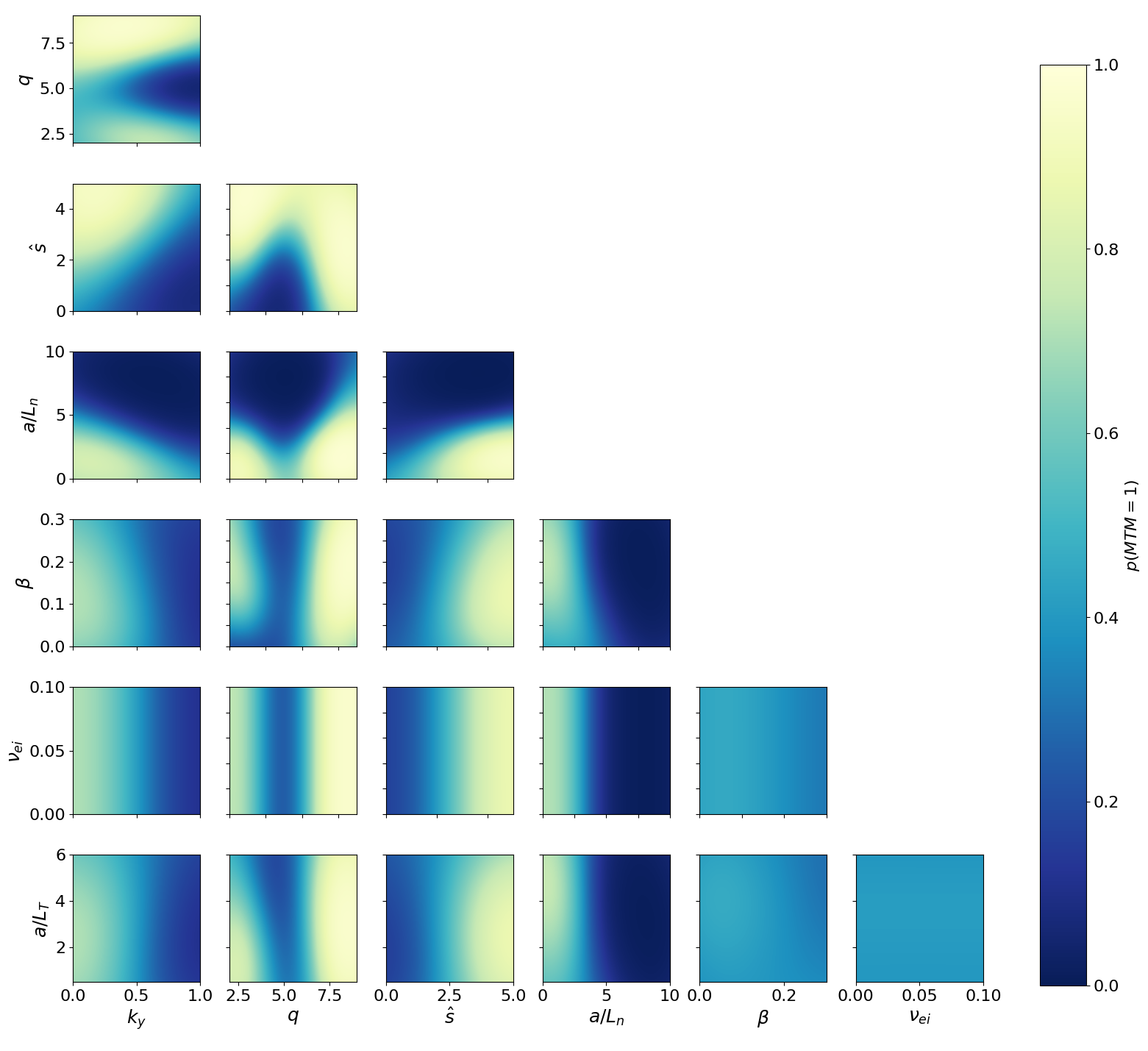}
    \caption{The probability of finding an MTM $p(\text{MTM} = 1|x)$ along planes in the input parameter space, according to the Bernoulli GP classifier with the best performing kernel function, namely Matérn 5/2. There is a complex structure across the parameter space showing broad areas of instability. The stabilising effect of high-density gradients is also apparent. For each marginal parameter scan, all remaining parameters were fixed at the centre point of their respective domain.}
    \label{fig:mtm_predictions}
\end{figure*}

Figure \ref{fig:mtm_predictions} shows the predictions of the Bernoulli GP Classifier with the Matérn 5/2 covariance kernel, which was the best performing BGP classifier, across each of the 2-dimensional subspaces of the 7-dimensional space. For each marginal parameter scan, all the remaining parameters were fixed at the centre point of their respective domain.  These indicate where in the parameter space MTMs are likely to be unstable.

\subsection{Regression}
GP Regression was employed to predict the mode frequencies, growth rates and normalised electron heat fluxes across the unstable regions of the parameter space using the simulations which were identified as MTMs as training data.  Independent GPs are trained for each output. Initial experiments in this included multi-task GP models (where correlations between outputs can yield more accuracy) with an Intrinsic Coregionalization Model (ICM) kernel on the outputs. However it was found that they did not yield any significant improvement for our data compared to the independent predictors and they are more expensive to train. 

Four different covariance functions (Matérn 1/2 (M12), Matérn 3/2 (M32), Matérn 5/2 (M52) and Radial Basis Function (RBF)) were investigated for each output variable. Each model used a constant mean function. Initial experiments additionally included corresponding models with a linear mean function, but this provided no predictive improvement. The results of a 5-fold cross-validation across the entire dataset of MTM-positive datapoints are shown in table \ref{tab:regression_results}. The MSLL metrics show that every GP model performs better than the trivial model $y_\star \sim \mathbb{N}(\mu_\mathcal{D}, \sigma_\mathcal{D}^2)$. The coverage shows that there is good agreement with the theoretical $95\%$ credible interval for every model. There is no appreciable difference between the kernels with respect to the chosen metrics. However, there is a trade-off between the predictive variance and the inferred observation noise with respect to the kernel smoothness (not shown). As the kernels become smoother, the uncertainty is pushed from the predictive variance to the observation noise. This is reflected in the calibration error, which decreases slightly but monotonically with the smoothness of the kernel (RBF being the smoothest and M12 being the roughest).

\begin{table}[htbp]
\centering
\footnotesize
\begin{adjustbox}{angle=90}
\begin{tabular}{lrrrrr}
\toprule
Model & MSE & SMSE & MSLL & CE & $C_{0.95}$ \\
\midrule
Growth Rate \\
\midrule
M12 & 0.069 ± 0.008 & 0.068 ± 0.016 & -1.325 ± 0.106 & 0.092 ± 0.011 & 0.952 ± 0.015 \\
M32 & 0.070 ± 0.008 & 0.062 ± 0.014 & -1.369 ± 0.116 & 0.083 ± 0.011 & 0.943 ± 0.017 \\
M52 & 0.070 ± 0.008 & 0.064 ± 0.015 & -1.363 ± 0.117 & 0.080 ± 0.011 & 0.939 ± 0.019 \\
RBF & 0.071 ± 0.008 & 0.066 ± 0.014 & -1.350 ± 0.107 & 0.081 ± 0.011 & 0.934 ± 0.020 \\
\midrule
Mode Frequency \\
\midrule
M12 & 3.338 ± 0.518 & 0.030 ± 0.011 & -1.692 ± 0.185 & 0.148 ± 0.011 & 0.957 ± 0.015 \\
M32 & 3.344 ± 0.517 & 0.029 ± 0.011 & -1.742 ± 0.187 & 0.138 ± 0.012 & 0.953 ± 0.019 \\
M52 & 3.336 ± 0.516 & 0.031 ± 0.012 & -1.722 ± 0.176 & 0.133 ± 0.012 & 0.948 ± 0.019 \\
RBF & 3.320 ± 0.504 & 0.038 ± 0.012 & -1.651 ± 0.154 & 0.122 ± 0.010 & 0.940 ± 0.024 \\								
\midrule
Quasilinear Flux \\
\midrule
M12 & 0.088 ± 0.014 & 0.039 ± 0.013 & -1.485 ± 0.130 & 0.165 ± 0.008 & 0.963 ± 0.018 \\
M32 & 0.088 ± 0.014 & 0.041 ± 0.013 & -1.493 ± 0.140 & 0.159 ± 0.009 & 0.960 ± 0.022 \\
M52 & 0.088 ± 0.014 & 0.043 ± 0.013 & -1.482 ± 0.138 & 0.156 ± 0.009 & 0.956 ± 0.023 \\
RBF & 0.088 ± 0.014 & 0.047 ± 0.013 & -1.451 ± 0.128 & 0.147 ± 0.008 & 0.952 ± 0.025 \\
\bottomrule
\end{tabular}
\end{adjustbox}
\caption{Mean Squared Error (MSE), Standardised MSE (SMSE), Mean Squared Log Loss (MSLL), Calibration Error (CE) and the coverage of the 95\% credible interval for each investigated model. Mean and standard deviation were computed from a 5-fold cross-validation. It should be noted that the MSE is not unitless and thus the values are significantly larger for the frequency than those for the growth rate and flux.  Unitless metrics such as the MSLL show comparable values between the three outputs.}
\label{tab:regression_results}
\end{table}

\begin{figure}
  \centering
  \includegraphics[width=\textwidth]{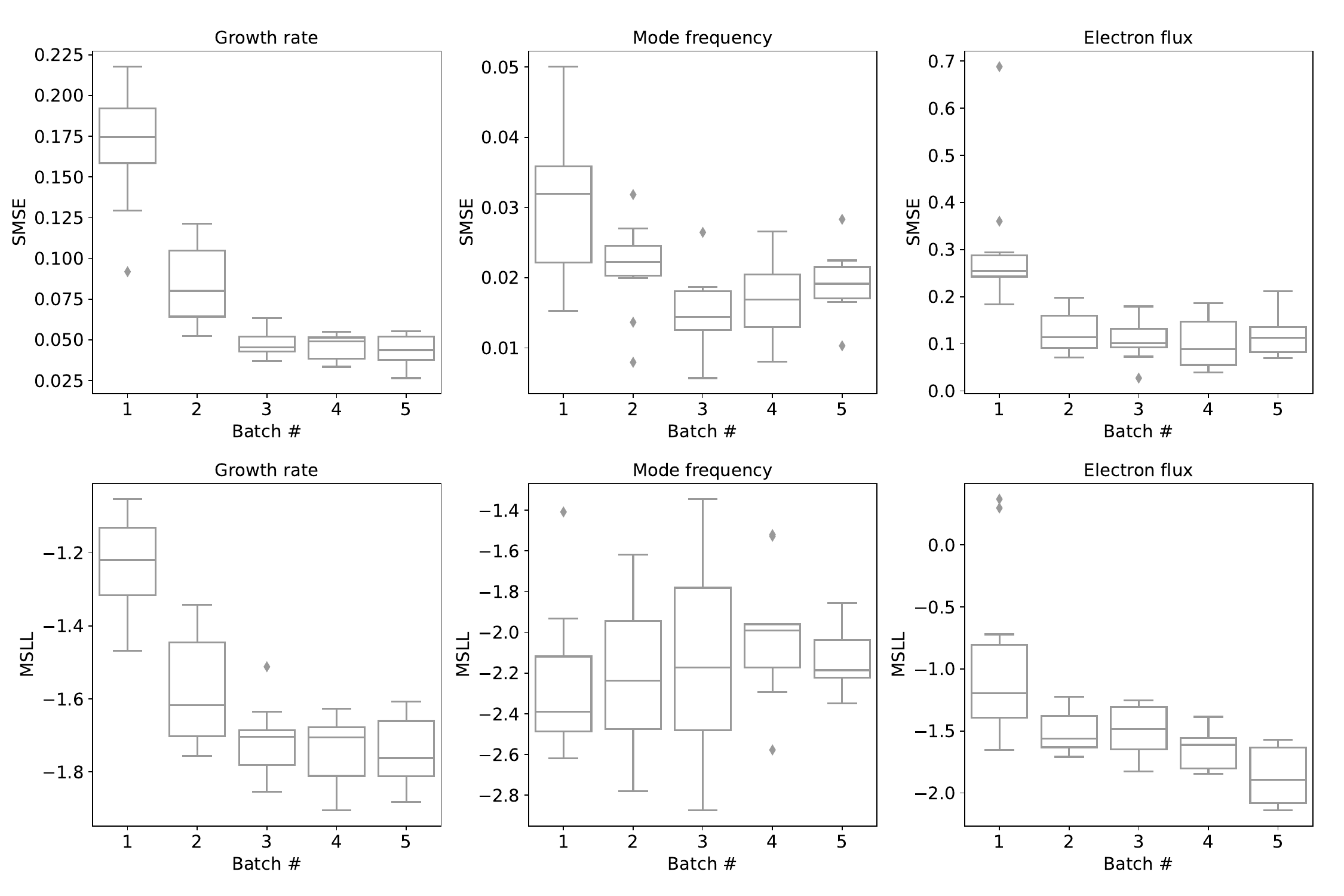}
    \caption{The evolution of the standardised mean squared error (SMSE) and the mean standardised log loss (MSLL) as more batches are added to the dataset, showing that for the growth rate and flux models, the predictive loss decreases as the data volume is increased. For the mode frequency models, the variance of the MSLL across batches first increases with batches 2 and 3 (for batch details see Table \ref{tab:batchstats}), and then decreases with batches 4 and 5. Note that batches 2 and 3  were targeted towards areas with a high probability of finding an MTM, while batches 4 and 5 were targeted towards areas with high flux, as explained in Table \ref{tab:batchstats}. For each boxplot, the box spans the first quartile (Q1) and the third quartile (Q3), while the centre line displays the second quartile (Q2), i.e. the median value. The whiskers show $1.5 \: \mathrm{IQR}$, where $\mathrm{IQR}$ is the interquartile range. The points show outliers.  The batch number correspond to the those described in Table. \ref{tab:batchstats}.}
    \label{fig:gpr_batch_development}
\end{figure}

\begin{figure}
  \centering
  \begin{subfigure}[t]{.3\linewidth}
    \centering\includegraphics[width=\linewidth]{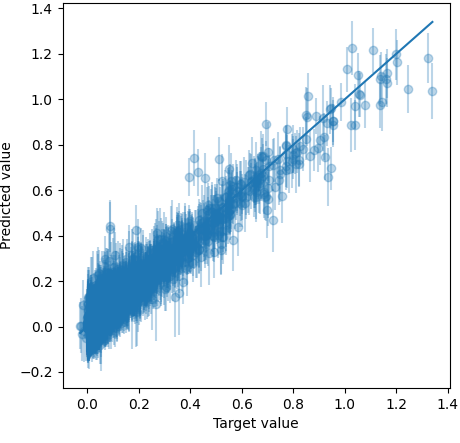}
  \end{subfigure}
  \begin{subfigure}[t]{.3\linewidth}
    \centering\includegraphics[width=\linewidth]{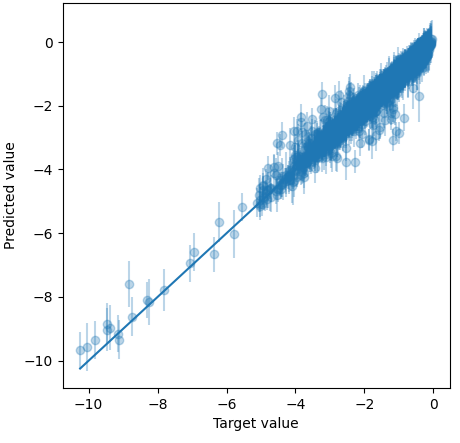}
  \end{subfigure}
  \begin{subfigure}[t]{.3\linewidth}
    \centering\includegraphics[width=\linewidth]{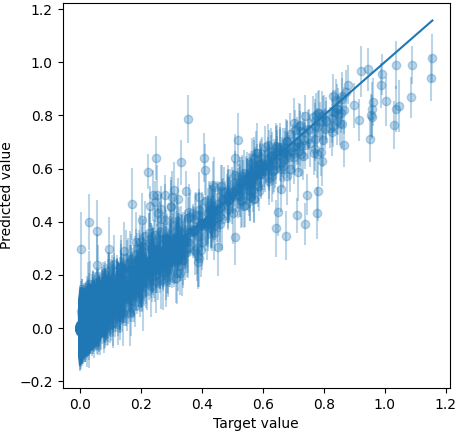}
  \end{subfigure} ~
  \begin{subfigure}[t]{.3\linewidth}
    \centering\includegraphics[width=\linewidth]{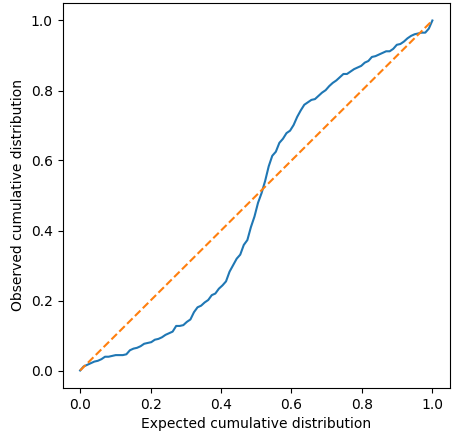}
    \caption{Growth Rate.}
  \end{subfigure}
  \begin{subfigure}[t]{.3\linewidth}
    \centering\includegraphics[width=\linewidth]{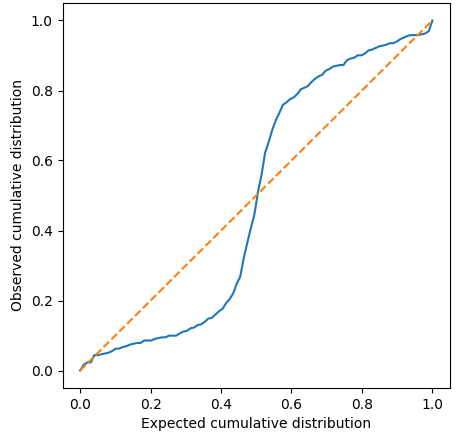}
    \caption{Mode Frequency.}
  \end{subfigure}
  \begin{subfigure}[t]{.3\linewidth}
    \centering\includegraphics[width=\linewidth]{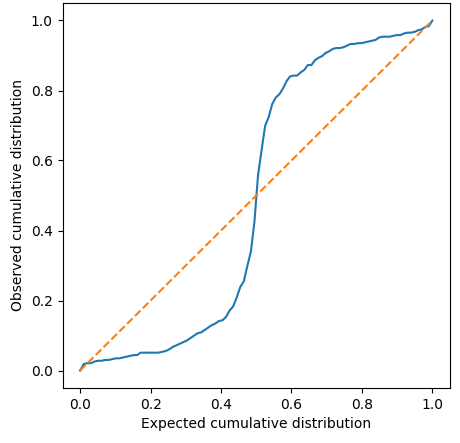}
    \caption{Electron Heat Flux.}
  \end{subfigure}
    \caption{Target vs predicted values (top row, error bars display the predictive standard deviation) and calibration curve $\nu(q)$ (bottom row) for the growth rate (left), mode frequency (centre) and quasilinear flux (right) for the models using the Matérn 1/2 covariance kernel.  An ideally calibrated model is shown as an orange line. Here a k-fold validation is performed ($k=5$, where 5 independent 80-20 splits of the data is performed to cover all the data.}
    \label{fig:gpr_test_performance}
\end{figure}

Figure \ref{fig:gpr_batch_development} shows the MSLL and SMSE of the GP model using the Matérn 1/2 covariance kernel, as more data points from the additional batches described in Section \ref{sec:activelearning} are cumulatively added.  The classifier and regression are retrained after each iteration.  Here, k-fold cross-validation with $k=10$ is performed, with the testing data for each fold kept fixed with $N_{test}=200$ data points from the first batch. Broadly, each new batch improves the performance of the growth rate and quasilinear heat flux predictors with respect to both metrics. However, for the mode frequency, the variance of the MSLL across the $k=10$ folds first increases and then decreases again. Notably, as the variance increases, some folds are better than any of the previous and some are worse. This shows there are some testing data that are highly sensitive to the new data points from MTM-targeted batches (batch 2 and 3, see Table \ref{tab:batchstats}). As the high-flux targeted batches (batches 4 and 5, see Table \ref{tab:batchstats}) are added, the variance of the MSLL contracts again, indicating that the sensitive test points are resolved. There is some correlation between the mode frequency and electron flux, which may explain why the high-flux batches assist in resolving the sensitive mode frequency data points. It is also worth noting that the MSLL of the mode frequency is consistently low.

\subsection{Individual parameter scans}
\label{sec:parameter}

\begin{figure*}
    \centering
    \includegraphics[width=1.0\textwidth]{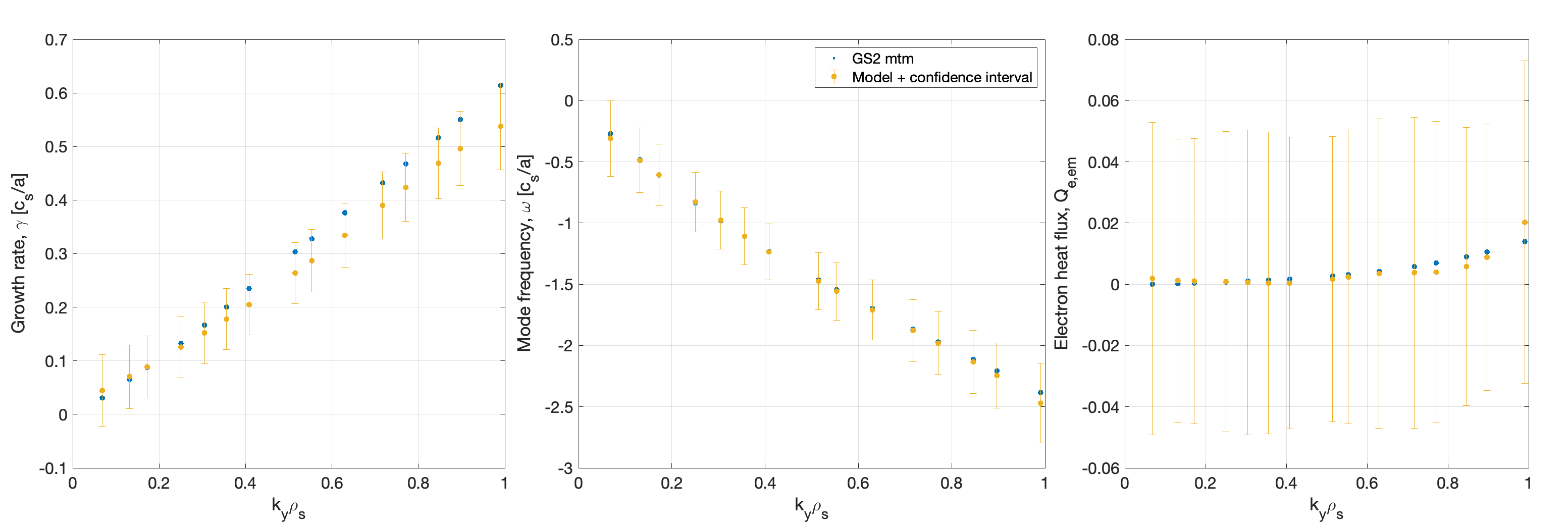}
    \caption{A parameter scan over $k_y$ away from marginal stability comparing the model output values (orange) with those predicted by GS2 (blue) for the growth rate (left), the mode frequency (middle) and the quasi-linear electron flux (right). The error bars represent the $95\%$ confidence interval as predicted by the GP regression. It should be noted that the GS2 simulations were not included in the training set and that the values all fall within the confidence intervals of the model. The other model parameters are set to: $q=2.43$, $\hat{s}=3.0$, $\beta=0.26$, $a/L_{ne}=0.29$, $a/L_{Te}=5.91$, $\nu_{ei}=0.054$.}
    \label{Fig:kyscanvalidation1}
\end{figure*}

\begin{figure*}
    \centering
    \includegraphics[width=1.0\textwidth]{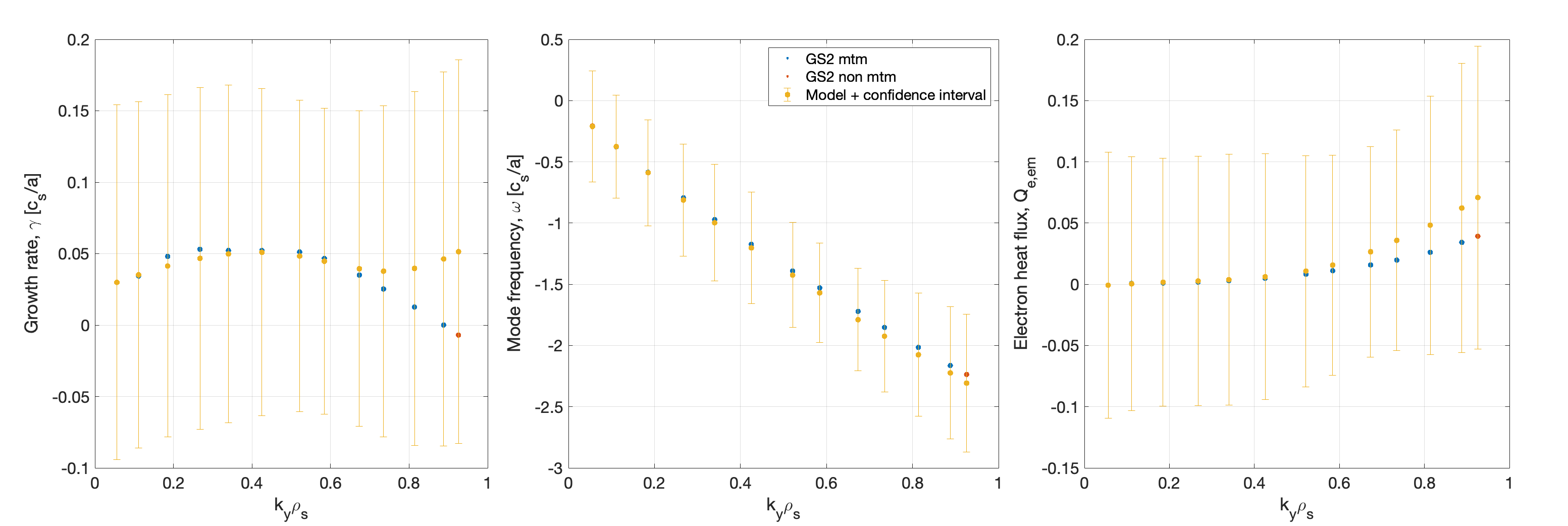}
    \caption{A parameter scan over $k_y$ close to marginal stability comparing the model output values (orange) with those predicted by GS2 (blue) for the growth rate (left), the mode frequency (middle) and the quasi-linear electron flux (right). The error bars represent the $95\%$ confidence interval as predicted by the GP regression. It should be noted that the GS2 simulations were not included in the training set and that the values all fall within the confidence intervals of the model. The other model parameters are set to: $q=4.45$, $\hat{s}=3.0$, $\beta=0.26$, $a/L_{ne}=0.18$, $a/L_{Te}=3.86$, $\nu_{ei}=0.094$.}
    \label{Fig:kyscanvalidation2}
\end{figure*}

The validation plots shown in Figure \ref{fig:gpr_test_performance}  showing the results of the k-fold validation of the whole dataset give an indication of the global performance of the model but do not show how well it can reproduce specific local parametric dependencies, which are important in integrated modelling. Figures \ref{Fig:kyscanvalidation1} and \ref{Fig:kyscanvalidation2} depict scans in the $k_{y}$ parameter comparing the predictions of the model and the values given by GS2. Two scans are shown, one far away from marginal stability (Figure \ref{Fig:kyscanvalidation1}) characterised by large growth rates, and one close to marginal stability (Figure \ref{Fig:kyscanvalidation2}), with lower growth rates and a characteristic peak in the growth rate spectrum. In both cases the model outputs are in good agreement with the GS2 predictions within the predictive uncertainties of the model.

The growth rate predictions (left, Figure \ref{Fig:kyscanvalidation2}) show two points of interest: Firstly the growth rate, and correspondingly the heat flux prediction, diverges from the actual values at the upper boundary of our $k_{y}$ range, indicating that there is a lack of data in this region requiring model improvement; secondly, the confidence intervals are significantly larger than the deviation of the model from the GS2 values, and more generally larger than the mean across this $k_{y}$ range indicating the complexity of capturing the zero growth rate manifold. The error bars here represent the \textit{predictive} standard deviation, i.e. $\sqrt{\mathbb V(f(x^\star)) + \sigma_\epsilon^2}$ where $\mathbb V(f(x^\star))$ is the variance of $f$ and $\sigma_\epsilon^2$ is the noise variance. The latter is practically zero and so the variance of $f$ dominates.

\begin{figure*}[h!]
    \centering
    \includegraphics[width=1.0\textwidth]{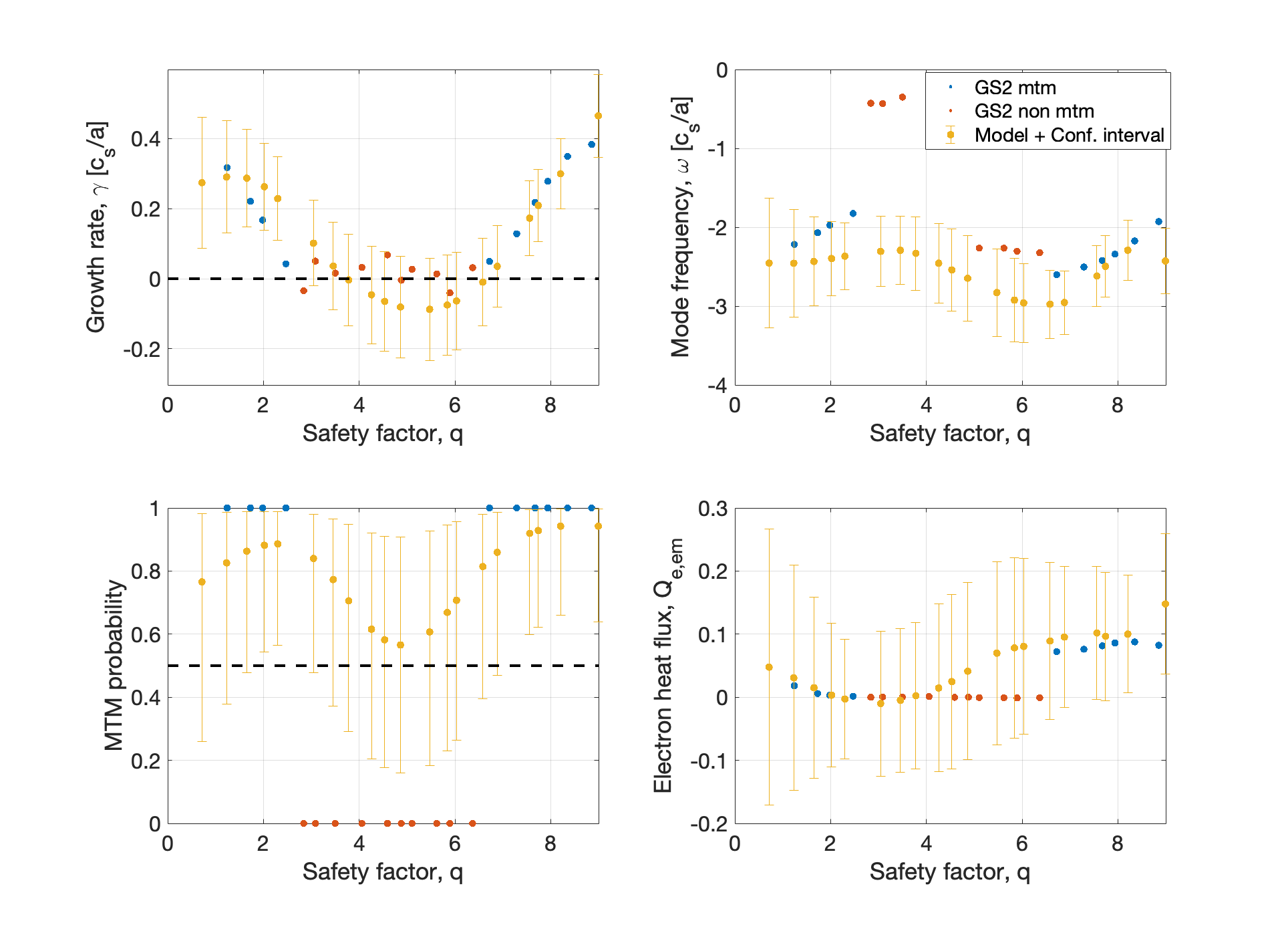}
    \caption{A scan over the safety factor $q$ comparing model output values (orange) with those predicted by GS2 (blue) for (top left) the growth rate, (top right) the mode frequency and (bottom right) the quasi-linear electron flux. The error bars represent the $95\%$ confidence interval as predicted by the GP regression. The other model parameters are set to: $q=4.45$, $\hat{s}=3.0$, $\beta=0.26$, $a/L_{ne}=0.18$, $a/L_{Te}=3.86$, $\nu_{ei}=0.094$. Black dashed lines indicate the zero growth rate line. (bottom left) (black) the classifier predicted MTM probability as a function of $\beta$. (blue) logical parameter denoting a GS2 run with an MTM with positive growth rate.}
    \label{Fig:qscanvalidation}
\end{figure*}

Figure \ref{Fig:qscanvalidation} shows a comparison of the model with GS2 output values over a scan in the safety factor, $q$. Here it is evident that there are two unstable branches in this part of parameter space, at low and high $q$. These two unstable branches are captured by the model, with the growth rate predicted to be negative at the values of q where GS2 does not predict an unstable MTM ($3 < q < 6$). This highlights the complex nature of the outputs, particularly the growth rate.  The model does not reproduce the frequency in between the two unstable branches well because this is a stable area where there is no data on which the GP model can train, thus the model reverts towards its mean.  Both branches maintain a dominant electromagnetic component ($|A_{||}|\geq |\phi|$) and a large value of the tearing parameter, thereby confirming that both branches are micro-tearing modes, however the mode structures in the higher branch are generally narrower in ballooning space, indicating a difference in which type of MTM the two branches contain.  An example of a lower $q$ and a higher $q$ can be seen in Figure \ref{fig:mtmexample}.

\begin{figure*}[h!]
    \centering
    \includegraphics[width=1.0\textwidth]{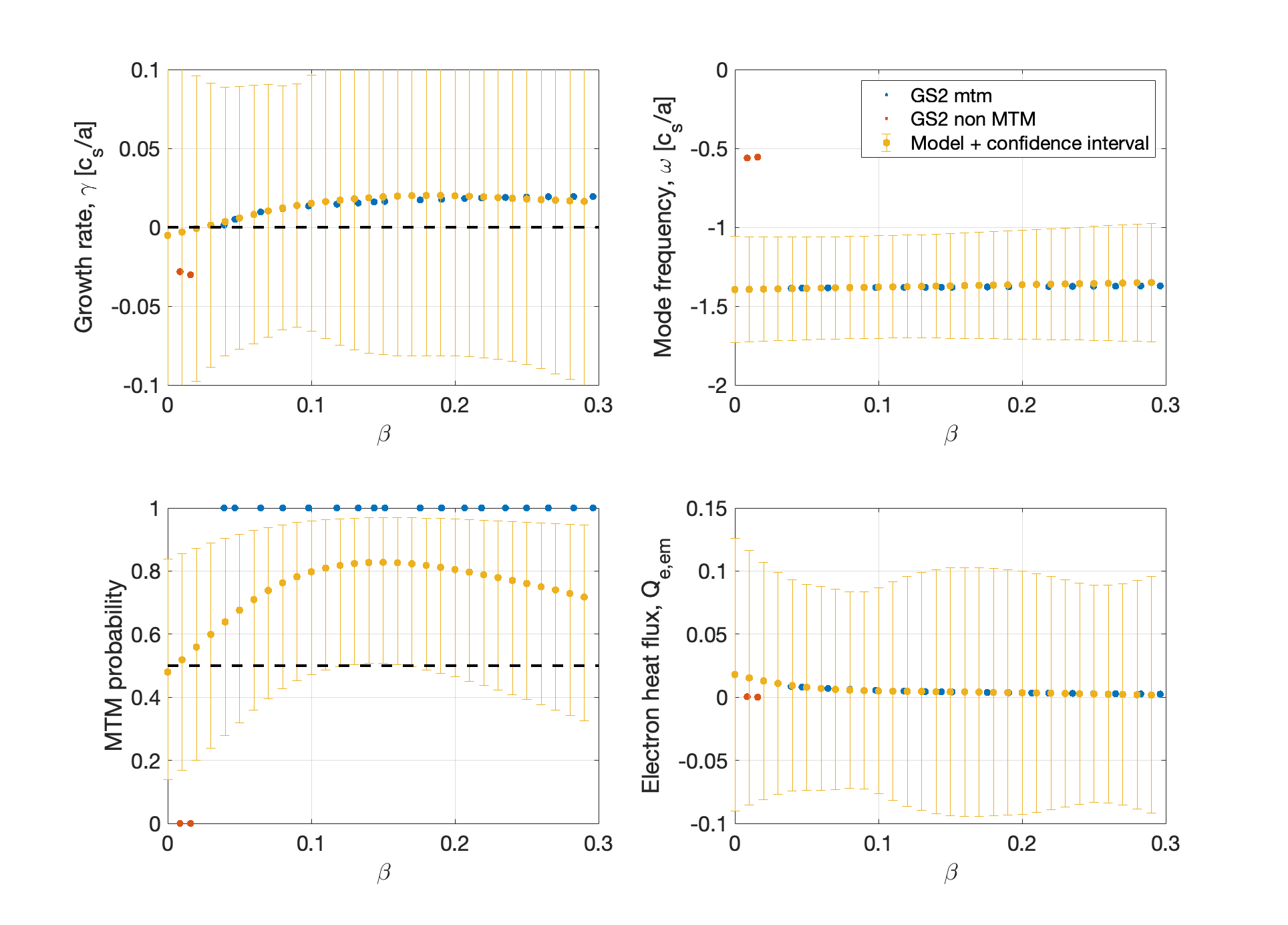}
    \caption{A scan over the plasma $\beta$ comparing model output values (orange) with those predicted by GS2 (blue) for (top left) the growth rate, (top right) the mode frequency and (bottom right) the quasi-linear electron flux. The error bars represent the $95\%$ confidence interval as predicted by the GP regression. The other model parameters are set to: $q=4.45$, $\hat{s}=3.0$, $k_{y}=0.26$, $a/L_{ne}=0.18$, $a/L_{Te}=3.86$, $\nu_{ei}=0.094$. Black dashed lines indicate the zero growth rate line. (bottom left) (black) the classifier predicted MTM probability as a function of $\beta$. (blue) logical parameter denoting a GS2 run with an MTM with positive growth rate.}
    \label{Fig:betavalidation}
\end{figure*}

A further crucial parameter that shows criticality is $\beta$. Figure \ref{Fig:betavalidation} shows that the model captures the stability boundary with high precision, again showing an agreement that is significantly better than the error bars suggest. In both figures \ref{Fig:qscanvalidation} and \ref{Fig:betavalidation} the classification of the data points in the scan and a comparison with the MTM probability as predicted by the GP based classifier show the expected dip in probability, below the 0.5 threshold at low $\beta$ and a corresponding dip between the branches within the confidence interval in the q-scan, which in both cases corresponds to the drop in the growth rate below 0 indicating a stable area.

\section{Conclusions and future development}
\label{sec:conclusions}

This paper described the development of Gaussian process-based models of the linear parameters of micro-tearing modes trained on data generated by a high fidelity gyrokinetic code.  Gaussian processes have advantages over neural network-based models for surrogate modelling via their ability to be more accurate at low data volume and their transparent and understandable confidence intervals.  

The model consists of two components: a classifier which predicts the probability of a given point in parameter space being unstable to MTMs, and a regressor which provides the quantity of interest in the unstable regions predicted by the classifier. Combining these in an active learning loop, it was possible to significantly increase the generation efficiency of training data on a previously unexplored parameter space.  An accurate model was created using approximately 5000 data points in a 7 dimensional parameter space requiring approximately 1M CPUhrs.  

A wider expansion of the model is planned, including:
\begin{itemize}
    \item Expansion of the parameter set to include all the geometric parameters listed in Table \ref{tab:fixedparams}. Increasing the dimensionality of the problem will require significantly expanding the data set.
    \item Harnessing the confidence intervals intrinsic in Gaussian processes to inform the targeting of new simulations in parts of parameter space where the most improvement can be made.
    \item Generalisation of the model to account for other types of mode which may be dominant, e.g KBM, hybrid KBM, trapped electron modes (TEM), etc.  This would benefit from the use of a gyrokinetic eigenvalue solver as this would allow the tracking of subdominant stable and unstable modes.
    \item Expanding the current parameters to cover larger ranges.  For example, it was apparent from figure \ref{Fig:kyscanvalidation1} that the current $k_{y}$ range is insufficient to resolve the turbulent spectrum peak in all cases.  It is thought that an expansion of this parameter to $k_{y}\rho_{s}\approx 7$ would be sufficient to resolve the peak.
    \item Further expanding the number of outputs to include the flux surface averaged $\langle k_{\perp}^{2} \rangle$.  A further component required for quasi-linear models.  This could be further refined to predict the full eigenfunctions of the perturbed fields.
    \item Integration of the model into the JINTRAC \cite{Jintrac2014} integrated modelling toolchain.  This will include the development of bespoke saturation rules required to predict the total electromagnetic heat flux via full nonlinear turbulence simulations.
\end{itemize}

\section*{Acknowledgements}
The authors would like to thank David Dickinson and the GS2 team for many discussions about the intricacies of the GS2 code.

\appendix

\section{Glossary of terms}
\label{sec:glossary}

When determining the accuracy of the GP classifier the following terms are used, they are listed here for the reader:

\begin{itemize}
    \item {\bf TP} - True positive - The fraction of points correctly classified as growing MTMs.
    \item {\bf FP} - False positive - The fraction of points incorrectly classified as a growing MTM.
    \item {\bf TN} - True negative - The fraction of points correctly classified as a stable point.
    \item {\bf FN} - False negative - The fraction of points incorrectly classified as a stable point
    \item {\bf Accuracy} - The fraction of true positive and true negative points - ${\bf( TP+TN)/(TP+TN+FP+FN)}$. 
    \item {\bf Precision} - The fraction of true positive over true positive and false positive - ${\bf TP/(TP+FP)}$.
    \item {\bf Recall} - The fraction of true positive over true positive and false negatives - ${\bf TP/(TP+FN)}$.
    \item {\bf F1 score} - A metric that accounts for both the precision and the recall, defined as the harmonic mean of the two, $F1 = 2*{\bf Precision}\times{\bf Recall}/({\bf Precision}+{\bf Recall})$
\end{itemize}

An ideal classifier would classify points as True positive or true negative only.  Multiple metrics are used to analyse a classifier in unbalanced systems (when number of MTM points are significantly more or less than non MTM points).  An ideal classifier would give values of one for its accuracy, precision, recall and F1.

\begin{itemize}
    \item {\bf MSE} - Mean squared error: $\text{MSE} = \frac{1}{N} \sum_{i=1}^N (y_i - f(x_i))^2$
    \item {\bf SMSE} - Standardised MSE: $\text{SMSE} = \frac{\text{MSE}}{\sigma_y^2}$. The MSE standardised by the variance of the test data.
    \item {\bf Coverage} - The coverage $C$ measures the coverage of the $n$th predictive credible interval with respect to the test data, with $n \in [0,1]$, i.e.
    \begin{equation}
        C_n = \frac{1}{N} \sum_{i=1}^N \mathbb I(y_i \in \text{CI}_n(x_i))
    \end{equation}
    where $\mathbb I(\cdot)$ is the indicator function and $\text{CI}_n(x)$ is the $n$ credible interval of testing datapoint $x$. For example, for the $0.95$ credible interval, the coverage should be $0.95$ for a well-calibrated model. Conversely, if $\text{CI}_n > n$, the model is underconfident with respect to that particular credible interval and vice versa.
\end{itemize}


\section*{References}
\bibliographystyle{unsrt}
\bibliography{bibliography}

\end{document}